\documentclass[journal=mamobx,manuscript=article]{achemso}
\usepackage{achemso}
\setkeys{acs}{usetitle = true}
\usepackage[version=3]{mhchem}
\usepackage[T1]{fontenc}
\usepackage{type1cm}
\usepackage{graphicx}
\usepackage{dcolumn}

\usepackage{bm}
\usepackage{units}
\usepackage{color}
\usepackage{latexsym}
\usepackage{subfig}
\usepackage{amsmath,amssymb}
\SectionNumbersOn

\title{The Role of the Functionality in the Branch Point Motion in Symmetric Star Polymers: A Combined Study by Simulations and Neutron Spin Echo}

\author{Stefan Holler}
\affiliation{Forschungszentrum J\"ulich GmbH, 52425 J\"ulich, Germany}
\alsoaffiliation{Centro de F\'isica de Materiales (CSIC, UPV/EHU) and Materials Physics Center MPC, Paseo Manuel de Lardizabal 5, 20018 San Sebasti\'an, Spain}

\author{Angel J. Moreno}
\affiliation{Centro de F\'isica de Materiales (CSIC, UPV/EHU) and Materials Physics Center MPC, Paseo Manuel de Lardizabal 5, 20018 San Sebasti\'an, Spain}
\alsoaffiliation{Donostia International Physics Center, Paseo Manuel de Lardizabal 4, 20018 San Sebastian, Spain}
\email{angeljose.moreno@ehu.es}

\author{Michaela Zamponi}
\affiliation{Forschungszentrum J\"ulich GmbH, 52425 J\"ulich, Germany}

\author{Petra Ba\v cov\'a}
\affiliation{Institute of Applied and Computational Mathematics (IACM), Foundation for Research and Technology Hellas (FORTH), 71110 Heraklion, Crete, Greece} 
\alsoaffiliation{Centro de F\'isica de Materiales (CSIC, UPV/EHU) and Materials Physics Center MPC, Paseo Manuel de Lardizabal 5, 20018 San Sebasti\'an, Spain}

\author{Lutz Willner}
\affiliation{Forschungszentrum J\"ulich GmbH, 52425 J\"ulich, Germany}

\author{Hermis Iatrou}
\affiliation{Department of Chemistry, University of Athens, Panepistimiopolis Zografou, 15771 Athens, Greece}

\author{Peter Falus}
\affiliation{Institut Laue-Langevin, 71 Avenue des Martyrs, CS 20156, 38042 Grenoble Cedex 9, France}

\author{Dieter Richter} 
\affiliation{Forschungszentrum J\"ulich GmbH, 52425 J\"ulich, Germany}

\begin{document}

\newpage

\begin{abstract}
We investigate the effect of the number of arms (functionality $f$) on the mobility of the branch point in symmetric star polymers.
For this purpose we carry out large-scale molecular dynamics simulations of simple bead-spring stars 
and neutron spin echo (NSE) spectroscopy experiments on center labeled polyethylene stars. 
This labeling scheme unique to neutron scattering allows us to directly observe the branch point motion on the molecular scale 
by measuring the dynamic structure factor.
We investigate the cases of different functionalities $f =$ 3, 4 and 5 for different arm lengths.
The analysis of the branch point fluctuations reveals a stronger localization with increasing functionality,
following $2/f$-scaling. 
The dynamic structure factors of the branch point are analyzed in terms 
of a modified version, incorporating dynamic tube dilution (DTD), of the Vilgis-Bou\'{e} model for cross-linked networks 
[J. Polym. Sci. B {\bf 1988}, {\it 26}, 2291-2302].
In DTD the tube parameters are renormalized
with the tube survival probability $\varphi (t)$.  As directly measured by the simulations, $\varphi (t)$ is independent of $f$ and therefore the theory
predicts no $f$-dependence of the relaxation of the branch point. 
The theory provides a good description of the NSE data and simulations for intermediate times. However, 
the simulations, which have access to much longer time scales, reveal the breakdown of the DTD prediction since
increasing the functionality actually leads to a slower relaxation of the branch point. 


\end{abstract}

\newpage
\maketitle

\section{Introduction}

Tube models \cite{Doi1988,McLeish2002,RubiColby,Graham2003,Read2008} are the standard approach for describing the viscoelasticity of entangled polymers. After the Rouse regime the chains feel the network of the topological constraints (`entanglements') originating from the uncrossability of the chains. This is represented as the motion of each chain in an effective tube (`reptation'). This framework, commonly accepted for linear chains, is changed in branched polymers. The presence of even a single branch point dramatically slows down the overall relaxation of the material
and leads to a complex rheological spectrum \cite{larson_branched,BoB,vanRuymbeke2006}. Solving this problem is relevant both from the basic and applied points of view due to the ubiquity of branching in industrial polymers \cite{Larson_book,science_daniel,Chen2012,Read2015}.

New relaxation mechanisms have been proposed for branched polymers as arm retraction
and branch point hopping \cite{Milner1997,Milner1998,Frischknecht2002}. According to arm retraction, 
the molecular segments in strongly entangled branched polymers relax hierarchically, progressing from the outer arm ends to the inner 
parts close to the branch points. This is a consequence of the reduced mobility of the branch point, which is chemically connected
to several arms that have no common tube to reptate in. Arm retraction is an exponentially slow process resulting in a broad distribution
of relaxation times along the arm \cite{Milner1997,Milner1998,McLeish2002,Bacova2013}. A direct consequence of this extreme dynamic disparity is that the relaxation of the inner segments is not hindered by the entanglements with the outer ones, which have relaxed at much earlier times.
This mechanism leads to constraint-release and is at the origin of dynamic tube dilution (DTD)
\cite{marrucci,mcleish_review_dtd}:
as the relaxation proceeds the inner segments  probe a tube that is progressively dilated over time.
Reptation remains inactive until the late stage of the relaxation \cite{Zhou2007,Bacova2014rs}, when all the side arms have relaxed and act as frictional beads for the effective linear chain \cite{larson_branched,Frischknecht2002,HierLarson,BoB,Kirkwood2009,Bacova2014}. In the case of symmetric architectures (e.g., star polymers with identical arms) reptation is fully suppressed, and branch point hopping is the single relaxation mechanism at late times \cite{BoB,HierLarson}.

The role of the functionality, i.e., the number $f$ of arms stemming from the branch point, has been scarcely explored in the literature. On the one hand, fluctuations of the branch point may be enhanced by increasing the functionality through more channels for `diving modes' \cite{Klein1986}, i.e. explorations of the tubes of the different arms. This may effectively broaden the original (`bare') tube for the branch point, as observed by Ba\v cov\'a {\it et al.} in simulations of symmetric 3-arm star polymers \cite{Bacova2013}. This `early tube dilution' (ETD) effect is fundamentally different from DTD, since diving modes do not produce constraint release \cite{Bacova2013}. On the other hand, 
diving modes involve some in-phase fluctuations of the arms around the branch point and these may be hindered by increasing the number of arms due to the generated drag and an increased number of possible chain interactions. In this case increasing functionality will lead to a stronger confinement  of the branch point. 
Identifying the correct mechanism (ETD vs confinement) is an important highlight to improve the tube models for branched polymers, which at present do not incorporate the potential role of the functionality.

In this article we expand, by combining molecular dynamics (MD) simulations and neutron spin echo (NSE) spectroscopy experiments,
previous investigations\cite{Zamponi2010,Bacova2013} of branch point motion in symmetric 3-arm star polymers to higher functionalities and different star sizes. We simulate a simple bead-spring model of melts of moderately entangled
($Z_{\rm a} =5$ entanglements per arm) symmetric stars of functionality $f = 3, 4$ and 5. 
We carry out NSE experiments in symmetric polyethylene stars of $f = 3$ and 4 arms. In the NSE experiments we also investigate
two different molecular weights, corresponding to $Z_{\rm a} = 5$ and 13 entanglements per arm. With a labelling scheme unique to neutron scattering, where only the nearest segments to the branch point are protonated, NSE allows us to directly observe the branch point motion on a molecular level through the measurement of the dynamic structure factor $S(q,t)$.

The analysis of the branch point fluctuations reveals a stronger localization of the branch point with increasing functionality,
in good agreement with the $2/f$-scaling proposed by Warner \cite{Warner1981}.
We discuss the simulation and experimental dynamic structure factors of the branch point in terms 
of a modified version of the model of Vilgis and Bou\'{e} \cite{Vilgis1988} for cross-linked networks where we incorporate DTD.
The MD and NSE data reveal that increasing the functionality leads to a slower relaxation of the branch point, including a different time dependence.
This result is not accounted for by current tube models and constitutes a violation of the expected
DTD renormalization of the tube parameters. The DTD theory indeed predicts no differences between stars with the same $Z_{\rm a}$ and different $f$, 
since the tube parameters are renormalized
with the tube survival probability $\varphi(t)$ and, as directly measured by the simulations,
$\varphi(t)$ is within statistics independent of the functionality.

The article is organized as follows. We present details of the MD simulation method in Section 2. Experimental details are given in Section 3.
The simulation and NSE results are analyzed and discussed in Sections 4 and 5 respectively. Conclusions are given in Section 6.

\section{Model and simulation details}
The simulated systems were based on the bead-spring model of Kremer and Grest \cite{Kremer1990}. The linear polymers and the arms of the stars were represented as linear chains of beads ('monomers') connected by springs. In each star all the arms were connected to the same central bead. In what follows the central bead of the star polymer or the linear chain will be denoted as the branch point. All the beads had the same diameter $\sigma$ and mass $m_0$. Excluded volume was implemented by introducing a repulsive Lennard-Jones (LJ) potential between the beads:

\begin{equation}
U_{\rm LJ}(r)=\begin{cases}
			4 \epsilon \left[\left(\frac{\sigma}{r}\right)^{12}-\left(\frac{\sigma}{r}\right)^6 + \frac{1}{4}\right], & \text{for} \; r \leq r_{\rm c}\\
			0, & \text{for} \; r > r_{\rm c}
			\end{cases}	 
\end{equation}	
with the cutoff at the distance $r_{\rm c} = 2^{1/6}\sigma$. The connecting springs were represented by a finitely extensible nonlinear elastic (FENE) potential \cite{Kremer1990}:
%
\begin{equation}
U_{\rm F} = -\frac{1}{2}K_{\rm F} R_{\rm F}^2 \ln \left[ 1-\left(\frac{r}{R_{\rm F}}\right)^2\right] 
\end{equation}
with a spring constant $K_{\rm F} = 30\epsilon/\sigma^2$ and a maximum bond length $R_{\rm F} = 1.5\sigma$. The LJ diameter $\sigma$ is the length unit of the model. We introduced an additional bending potential between every three consecutive beads in order to implement local stiffness. The bending potential was given by:
%
\begin{equation}
U_{\rm bend}(\theta) = a_2\theta^2 + a_4 \theta^4
\end{equation}
where $\theta$ is the angle between consecutive bond vectors.  We used the values $a_2/\epsilon= 0.92504230$ and $a_4/\epsilon = -0.054183683$, which in practice produce a potential undistinguishable from the cosine potential used in Ref.~\cite{Bacova2013} (the differences are negligible except for large angles that are never accessed due to the repulsive LJ interaction between bonded beads). The used bending potential provides a slightly semiflexible character to the chains (characteristic ratio \cite{Puetz1999,Bacova2013} $C_{\infty} \approx 3.6$ vs. $C_{\infty} \approx 1.8$ for the flexible case $a_2 = a_4 = 0$), and lowers the entanglement length. Thus, an analysis of the primitive path \cite{Everaers2004,Bacova2017} for this semiflexible model at the same simulated density gives an entanglement length of $N_{\rm e} \approx 25$ monomers, whereas the same analysis gives $N_{\rm e} \approx 65 - 85$ for the flexible case \cite{Everaers2004,Hoy2009}.  This allows for simulating more strongly entangled systems than in the flexible case by using the same chain length.

\begin{figure}[!ht]
{\centering \resizebox*{0.5\textwidth}{2.5in}{\includegraphics{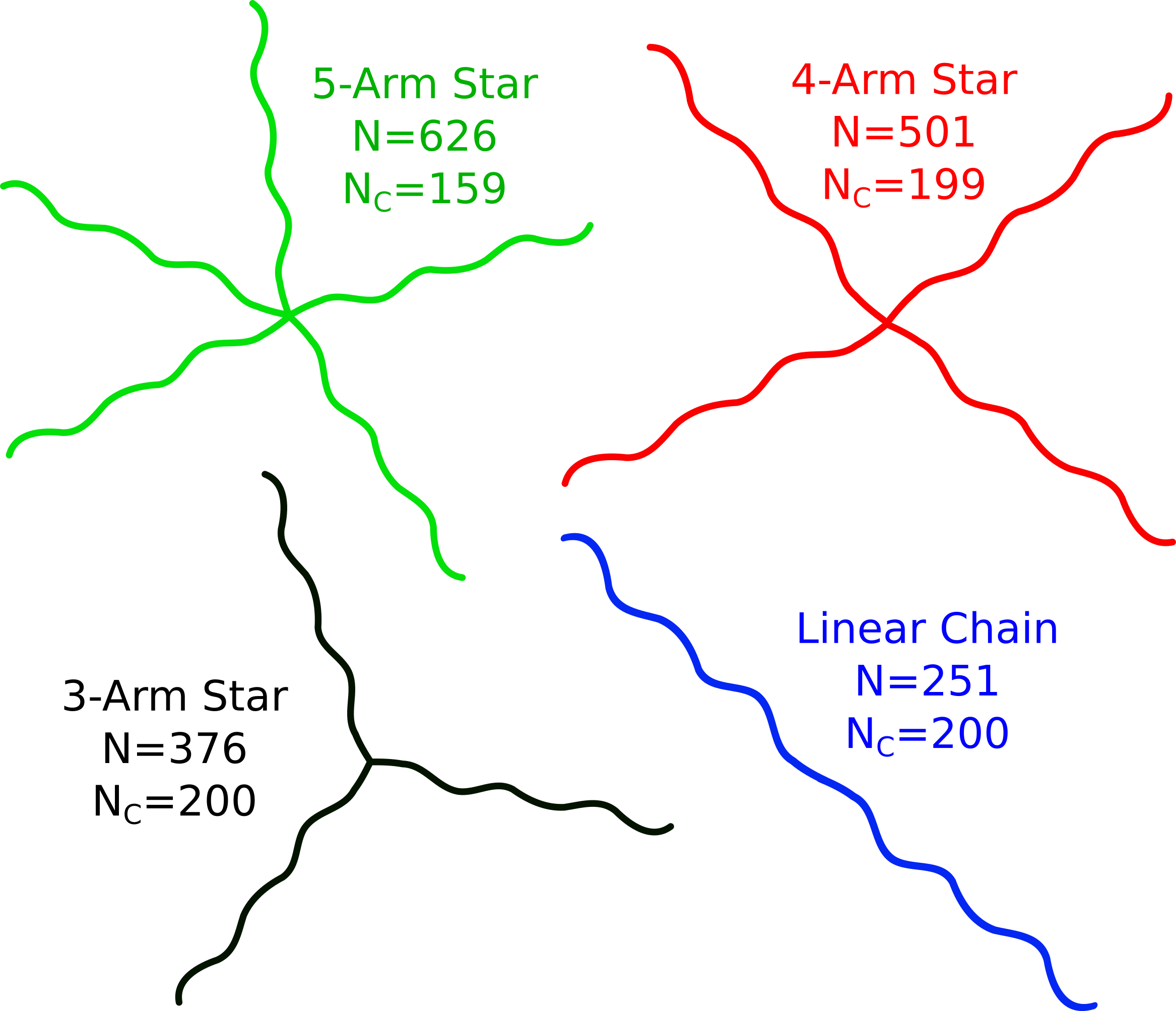}} \par}
\caption{ Schematic representation of the simulated systems. $N_{\rm C}$ is the number of polymers in the simulated box. $N$ is the number of beads per polymer. Each arm is 125 beads long (corresponding to $Z_{\rm a} \approx 5$ entanglements).}
\label{simu-geom}
\end{figure}

In what follows length and time in the simulations will be given in units of $\sigma$ and $\tau_0 = (m_0 \sigma^2/\epsilon)^{1/2}$, respectively. The simulations were 
performed at temperature $T = \epsilon/k_{\rm B}$ (with $k_{\rm B}$ the Boltzmann constant) 
and at monomer density 
$\rho = 0.85\sigma^{-3}$, which qualitatively corresponds to melt density \cite{Kremer1990}.  We simulated three systems of stars with $f$ = 3, 4 and 5 arms and a system of linear chains (the central bead and two `arms', $f$ = 2). The systems were monodisperse, i.e., all the polymers in the same system were identical in the number of arms $f$ and the number of monomers per arm, $N_{\rm a}$. We used $N_{\rm a} = 125$, 
which corresponds to $Z_{\rm a} = N_{\rm a}/N_{\rm e} \approx 5$ entanglements per arm (so that the backbone of the linear chain had 251 monomers and $Z \approx 10$ entanglements). A schematic representation of the four simulated architectures is shown in \ref{simu-geom}. $N$ is the total number of beads in the star and $N_{\rm c}$ is the number of polymers in the cubic simulation box, which contained a total number of monomers ranging from 50200 (linear chains) to 99534 (5-arm stars). Periodic boundary conditions were implemented. 

The systems were generated and equilibrated following the procedure of Refs.~\cite{Auhl2003,Bacova2013}. The stars and the linear chains were first constructed with the correct statistics for the intramolecular distances. This was obtained with low computational cost by simulations of weakly entangled chains, which reached the Gaussian regime at relatively short contour distances \cite{Auhl2003,Bacova2013}. Then the desired architecture was created by joining the weakly entangled parts and the obtained polymer structures were placed at random positions in the simulation box and we followed the prepacking procedure proposed by Auhl {\it et al.} \cite{Auhl2003}. This consisted of a Monte Carlo simulation where the polymers were treated as rigid objects performing large-scale motions (translations, rotations, reflections, etc), which were accepted only when local density fluctuations were reduced. After a significant reduction of the inhomogeneities, the last step of the equilibration consisted of a MD run in which force-capping was applied to the LJ interaction. The capping radius was progressively reduced and finally the full LJ potential was switched on. Before starting the production runs we verified that the polymers had recovered the correct statistical properties. Further details of the equilibration procedure can be found in Ref.~\cite{Bacova2013}.

%

The longest production runs extended up to $4 \times 10^8$ time steps. We also performed MD simulations with fixed end monomers of the star arms. This allowed us to investigate the systems in the absence of the constraint release events associated to arm retraction, which is obviously suppressed when the arm ends are fixed.  The MD simulations were performed with GROMACS \cite{Hess2008}. Langevin dynamics were used with a friction $\gamma = 0.5 m_0/\tau_0$ and a time step $\Delta t = 0.005\tau_0$.

\section{Experimental details}

\subsection{Star Polymer Synthesis and Sample Preparation}

For the neutron spin echo experiments we have prepared two 3-arm and two 4-arm center-labeled polyethylene (PE) star polymers. The star arms are effectively diblock copolymers consisting of a relatively long deuterated and a short protonated PE block. The arms are covalently attached to a central branch point via the protonated block such that the inner part of the star is labeled. Consistent with previous studies \cite{Zamponi2010}, the label  size per arm was  kept constant for all stars at  about 1 kg/mol, which corresponds to approximately  1/2 of the entanglement mass. In order to keep consistency between the scattering of the 3- and 4-arm stars, only three arms of the 4-arm stars were labeled, whereas the remaining 4th arm was fully deuterated. For each star functionality  we have prepared a `small' and a `large' version. The arms of the small stars have an overall molecular weight  of  9 kg/mol, which corresponds to about 5 entanglements. The arms of the large stars were approximately 26 kg/mol corresponding to 13 entanglements. Due to the comparatively small protonated centers, the majority component is deuterated PE acting as a kind of invisible matrix in the NSE experiment. In this way, a direct access to the branch point motions became  possible. \\ 

\noindent The star polymers were prepared by a three step synthesis. In the first step, single arms were synthesized by living anionic polymerization of 1,3-butadiene. The living arms were subsequently linked in a second step to methyltrichlorosilane CH$_3$SiCl$_3$ and tetrachlorosilane SiCl$_4$, respectively, to produce 3-arm and 4-arm polybutadiene stars. In the third step, the polybutadiene stars were saturated with deuterium by means of a palladium catalyst leading to well-defined, temperature stable polyethylene stars. 

The polymerizations of the PB-arms and the linking reactions were carried under high vacuum using custom made glass reactors equipped with  break seals for the addition of reagents. Details of this technique including description of reactors, preparation of initiator, purification of monomer, solvents and chlorosilanes have been published earlier in Ref \cite{Hadjichristidis2000}. 
Briefly, labeled PB-d$_6$-PB-h$_6$ arms with 9 (8-1) kg/mol and 26 (25-1) kg/mol overall molecular weight were synthesized by sequential addition of 1,3-butadiene-d$_6$ (98.9 \%D, Cambridge Isotope Laboratories) and  1,3 butadiene-h$-6$, respectively. A larger batch of the PB-d$_6$-PB-h$_6$ arms  with 9 (8-1) kg/mol was made and split to prepare  the small 3- and 4-arm stars with identical arms and label sizes. The PB-d$_6$-PB-h$_6$ arms  with  26 (25-1) kg/mol were prepared separately for each star. The reason for that was that the large 3-arm star was already synthesized earlier for a previous study by Zamponi  {\it et al.} \cite{Zamponi2010}. The single fully deuterated PB-d$_6$ arm with 9 kg/mol for the small 4-arm star was prepared separately, while the one with 25 kg/mol is identical to the first block of PB-d$_6$-PB-h$_6$ with 26 (25-1) kg/mol. In all cases t-butyllithium was taken as initiator and benzene  as polymerization solvent leading to  polybutadienes with a random composition of 93\% 1,4- and 7\% 1,2 addition. Small aliquots of each PB arm and block  were  removed prior to the linking reaction for separate characterization.

The synthesis of the 3-arm stars was then accomplished by reacting a small excess of the  living arms with methyltrichlorosilane. The reaction mixture was kept for one  month to ensure complete functionalization. The small amount of residual living arms was subsequently terminated with methanol and removed by fractionation using toluene/methanol as solvent/non-solvent pair.
 
For the preparation of the 4-arm stars at first the deuterated polybutadiene arm without protonated label was attached to tetrachlorosilane. This was accomplished by reacting with a large excess of SiCl$_4$ in order to avoid multiple substitution. The excess tetrachlorosilane was removed by distilling off all volatile components including the solvent. The residual dry d-PB-SiCl$_3$ was redissolved in dry benzene and reacted with a small excess of the living PB-d$_6$-PB-h$_6$ labeled arms. As described before the linking reaction was kept for 1 month, excess living arms were terminated and finally removed by fractionation.
 
Star polymers, arms and blocks were characterized by size exclusion chromatography (SEC). As chromatographic instrument we have used Infinity 1260 components (isocratic pump (G1310B) and auto sampler (G1329B)) from Agilent Technology combined with a column oven (CTO-20AC) from Shimadzu and a detection system from Wyatt Technology consisting of a differential refractive index concentration detector (Optilab rex) and a multi angle laser light scattering detector (Heleos Dawn 8$^+$) for absolute molecular weight determination. SEC data were obtained with tetrahydrofuran as eluent at $30^\circ\text{C}$ using three Agilent PolyPore GPC columns with a continuous pore size distribution at a flux rate of 1 mL/min. The chromatograms of the individual arms show  single narrow peaks indicating almost monodisperse molecular weight distributions. Chromatograms of the star polymers show in all cases a shallow signal beside the main peak corresponding to a small residual amount (< 1\%) of single arms. Since we do not expect  significant effects from the presence of small amount of single chains on the experimental results  we resigned from further fractionations which further keeps the yield  sufficiently high. A summary of exact molecular weights for the star polymers and arms are given in \ref{tbl:samples}.
 
\begin{table}
  \caption{Molecular Weight Characteristics of 3-arm and 4-arm polybutadiene star polymers}
  \label{tbl:samples}
  \begin{tabular}{llllll}
    \hline
    star & $M_w^{a)}$ 	& $M_w$(PB-d$_6$) & $M_w$ (PB-d$_6$-PB-h$_6$) & $M_w$ (single PB-d$_6$) & $M_w / M_n $ \\
     	 & [kg/mol] 		& [kg/mol] 		  & [kg/mol] 				 & [kg/mol] 				  & (star) 		\\
    \hline
    3-arm small  &	27.3 &  $ 8.2$ 		&   9.2 		& - 		 & 1.02	\\
    4-arm small	 &	35.7 &  $ 8.2$ 		&    9.2 	& 8.8 	 & 1.02	\\
    3-arm large  &	74.3	 &  $ 25.8 $		&  26.85 	&  - 	 & 1.04	\\ 
    4-arm large	 &	105 	 &  $ 24.8 $ 	& $25.8 $	& $24.8$ & 1.04	\\
    \hline
  \end{tabular}
 \\ $^{a)}$ Weight average molecular weight  determined by multi detector SEC.
 \end{table}

In the final step of the synthesis the polybutadiene stars were saturated with deuterium. As a catalyst we have used palladium on barium sulfate (5\% Pd, unreduced). The saturation was done in a 1L stainless steel, high pressure  reactor (Berghof) in cyclohexane at $90^\circ\text{C}$ and 45 bars under rigorous stirring. After the reaction the catalyst was removed by filtration at elevated temperature ($70^\circ\text{C}$). The final PE stars were obtained by precipitation in a mixture of acetone and methanol (3:1 v/v). The labeled center of the final PE stars have a nominal d,h composition of d2h6. Earlier NMR experiments have shown that in addition to the saturation a small extent of d-h substitution takes place under the applied conditions leading to a decreased h-content of the label of approximately h5d3. We have however not  determined the exact h,d-composition of the synthesized stars as it is irrelevant for the studies presented in this work.
 
For the NSE measurements the PE stars were filled into Niobium containers. Homogeneously filled specimen were obtained by repeatedly filling and melting the crystalline PE in a vacuum oven at $100^\circ\text{C}$. Finally, the containers were sealed with teflon and closed in an argon glove box to avoid oxidation at the measuring temperature of 509 K.

\subsection{Neutron spin echo spectroscopy}

NSE spectroscopy measures directly the intermediate scattering function\cite{Mezei1980} $S(q,t)$, given by:
\begin{equation}
S_{\rm coh}(q,t)=  \sum_{m,n} \left< \exp(iq \cdot [\vec{r}_m(t)-\vec{r}_n(0)])\right>
\label{eq:s-q-t}
\end{equation}
with $\vec{r}_m$ the position of the monomer $m$.
The momentum transfer $q$ is related to the wavelength $\lambda$ in the quasielastic case as $q = (4\pi/\lambda) \sin(\theta/2)$ with $\theta$ being the scattering angle. The neutron scattering length of deuterons and protons is very different ($b_{\rm coh}$ in fm: H = -3.74 and D = 6.67), which leads to a contrast between the labeled part and the deuterated matrix. We use specific selective labeled stars with a small amount (compared to the deuterated matrix) of label at the branch point so we can measure the coherent dynamic structure factor that is calculated using the Gaussian approximation as described later in \ref{eq:gaussian}. This shows the internal segment-segment pair correlations of the labeled section.
The measurements were performed at the IN15 at the Institute Laue-Langevin, at a temperature of 509 K and wavelengths of $10 \text{\AA}$ and $14 \text{\AA}$. We used $q$ values from 0.05 to 0.115 $\text{\AA}^{-1}$. The background was measured with fully deuterated linear polyethylene chains and we corrected for it and the instrument resolution.

\section{Simulation results}

\subsection{Real-space dynamics}

\begin{figure}[!ht]
{\centering \resizebox*{0.5\textwidth}{2.5in}{\includegraphics{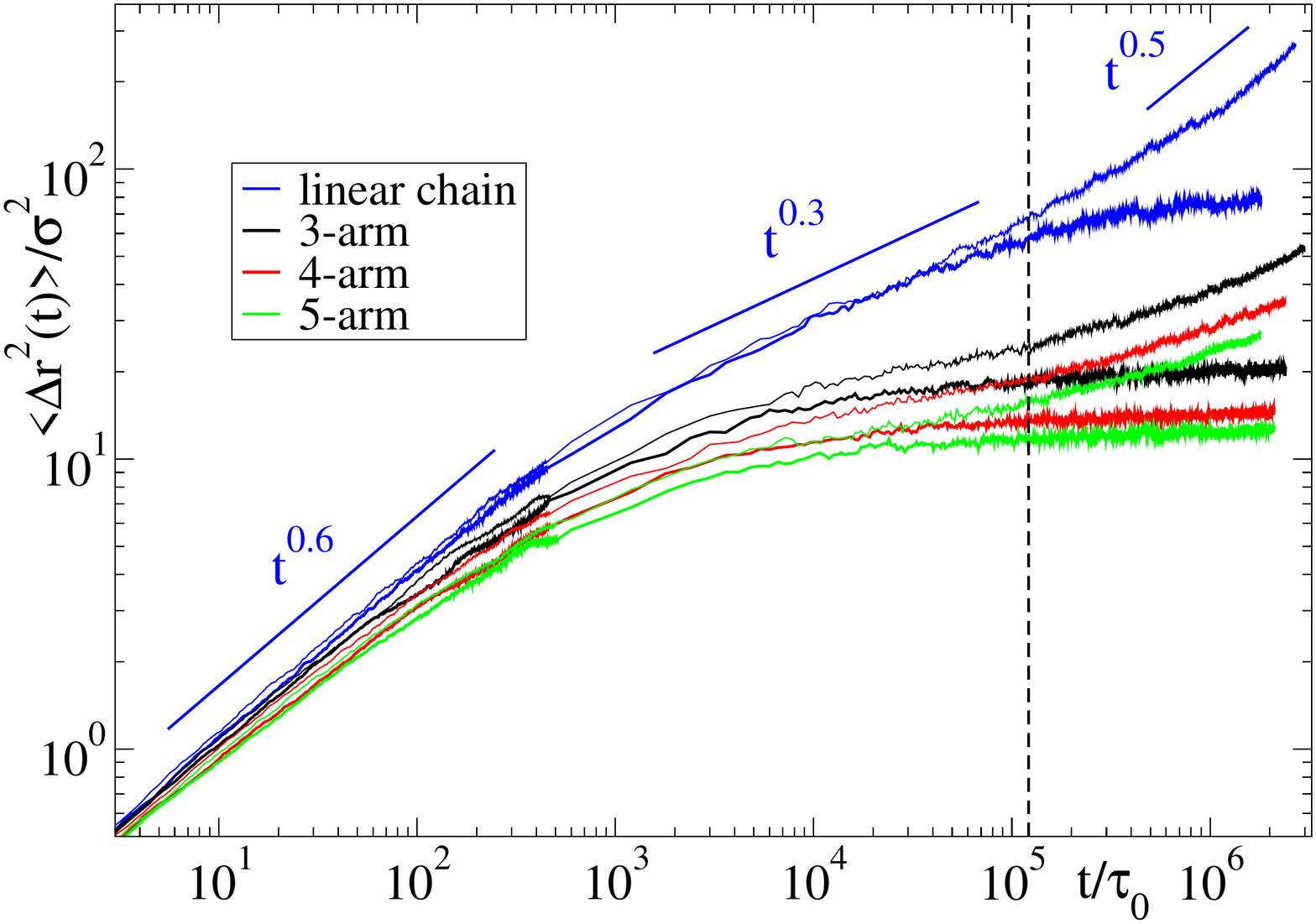}} \par}
\caption{Simulation results for the MSD of the molecular center (branch point and three nearest beads in each arm). Data are shown both for the simulations with free and fixed ends (the latter displaying the long-time plateau).  Blue, black, red and green curves correspond to the systems with functionality $f = 2$ (linear chains) and $f$ = 3, 4 and 5, respectively. The solid lines indicate, from short to long times, the different power-law regimes $\langle \Delta r^2 \rangle \sim t^{0.6}$ (Rouse), $t^{0.3}$ (Rouse in tube) and $t^{0.5}$ (reptation) in the linear chains.
The vertical dashed line indicates the time scale equivalent to the limit of the NSE window, $t_{\rm NSE} = 400$ ns (see below).}
\label{msd}
\end{figure}

\ref{msd} shows the mean square displacement (MSD) of the molecular centers in the four investigated systems. In order to improve statistics, we represent the MSD of the group of beads formed by the branch point and its three nearest beads in each arm. \ref{msd} includes the results from both the simulations with free and fixed ends. Regarding the simulations with free ends, the tube theory predicts the following power-law regimes for the MSD of the linear chains \cite{Doi1988,McLeish2002,RubiColby,Likhtman2002}:
%
%
\begin{equation}
\left<r^2\right> = \begin{cases}
		t^{\frac{1}{2}} 	\hspace{1cm} \text{for}	\hspace{1cm} \tau_{\rm s} < t < \tau_{\rm e} & \text{Rouse}	\\
		t^{\frac{1}{4}}\hspace{1cm}	\text{for}	\hspace{1cm} \tau_{\rm e} < t < \tau_{\rm R} & \text{Rouse in tube}	\\
		t^{\frac{1}{2}} 	\hspace{1cm} \text{for}	\hspace{1cm} \tau_{\rm R} < t < \tau_{\rm d} & \text{reptation}	\\
		t				\hspace{1.3cm} \text{for} \hspace{1cm}\tau_d < t 			& \text{diffusion}
\end{cases}
\end{equation}
where $\tau_{\rm s}$,$\tau_{\rm e}$,$\tau_{\rm R}$ and $\tau_{\rm d}$ are the monomeric, entanglement, Rouse and disentanglement times, respectively. From the monomeric time scale ($\tau_{\rm s} \approx \tau_0$) to the entanglement time ($\tau_{\rm e} \approx 1800\tau_0$ \cite{Bacova2013}) the simulated linear chains show an exponent 0.6 instead of the ideal Rouse value 0.5. In the Rouse-in-tube regime, from $\tau_{\rm e}$ to the Rouse time of the linear chains \cite{Doi1988} $\tau_{\rm R} = \tau_{\rm e} Z^2 \approx 1.8 \times 10^5 \tau_0$, an exponent 0.3 is found instead of the ideal value 0.25. These deviations may originate from the slightly semi-flexible character of the chains, which leads to short-range non-Gaussian correlations that are not included in the Rouse model. After the Rouse time the MSD of the linear chains is consistent with the expected reptation behavior $\langle \Delta r^2 \rangle \sim t^{1/2}$. The disentanglement time of the linear chains is expected at \cite{Doi1988} $\tau_{\rm d} = 3\tau_{\rm e}Z^3 \approx 5.4 \times 10^6 \tau_0$. This is beyond the simulated time and the final transition to diffusion is not reached.


In comparison to the linear chains, the molecular centers of the stars show a slower relaxation after the Rouse regime. This effect becomes more pronounced the higher the functionality. By fixing the arm ends DTD does not occur, since arm retraction and the associated constraint-release events are suppresed. In such conditions the branch point is expected to be localized. This is confirmed by the plateau in the MSD observed at $t > 10^4 \tau_0$
for the MSD of the systems with fixed ends. However a closer inspection of the data reveals that the plateau is not fully horizontal. The MSD still shows a marginal increase with time, suggesting the presence of some minor relaxation mechanism even if the arm ends are fixed. We speculate that end-looping events \cite{Zhou2006} might be non-negligible for the moderately entangled arms ($Z_{\rm a} \approx 5$) of the investigated systems. 

The plateau heights in the MSDs of the simulations with fixed ends accounts for the fluctuations of the branch point within the local tube. We obtained them by calculating the mean value for times $t > 4 \times 10^5 \tau_0$. The corresponding values are given in \ref{tbl:plateau-height}. According to Warner's theory \cite{Warner1981} the fluctuation within the tube measured by the plateau height scales as $\langle r^2 \rangle = p_f  \sim 2/f$, with $p_f$ the plateau height of the system with functionality $f$ and fixed ends. We test this prediction by computing the different ratios ($p_f /p_f'$) between plateau heights of systems with different functionalities $f, f'$, and the theoretically expected values $f'/f$  (see \ref{tbl:plateau-ratio}). A good agreement with theory is found for all the pairs of stars ($3 \leq f \leq 5$).  
On passing, we mention that if the ratio involves linear chains ($f = 2$), the simulation values deviate by a factor about 2.5 from the blind analysis based on Warner's theory. This is not suprising, since the theory is designed for points that are explicitly cross-linked \cite{Warner1981}.


\begin{table}
  \caption{Plateau height of the MSD in the simulations with fixed ends}
  \label{tbl:plateau-height}
  \begin{tabular}{ll}
    \hline
      			System & Plateau height ($\sigma^2$) \\
    \hline
			linear chain & 	$75.3 \pm 3$ \\
			3-arm star	&	$20.1 \pm 0.5$ \\
			4-arm star	&	$14.3 \pm 0.3$ \\
			5-arm star	&	$12.4 \pm 0.3$ \\						
    \hline
  \end{tabular}
\end{table}

\begin{table}
  \caption{Simulation (from \ref{tbl:plateau-height}) and theoretical values of the ratios of the MSD plateau heights in the systems with fixed ends. The plateau height is denoted as $p_f$ for the system with functionality $f$.}
  \label{tbl:plateau-ratio}
  \begin{tabular}{llll}
    \hline
      			 & $p_5/p_4$ & $p_5/p_3$ & $p_4/p_3$  \\
    \hline
			theory & 	1.25 & 1.66 & 1.33 \\
			simulation	&	$1.15 \pm 0.05$ &$1.62 \pm 0.08$ & $1.40 \pm 0.07$\\
						
    \hline
  \end{tabular}
\end{table}

\begin{figure}[!ht]
{\centering \resizebox*{0.5\textwidth}{2.5in}{\includegraphics{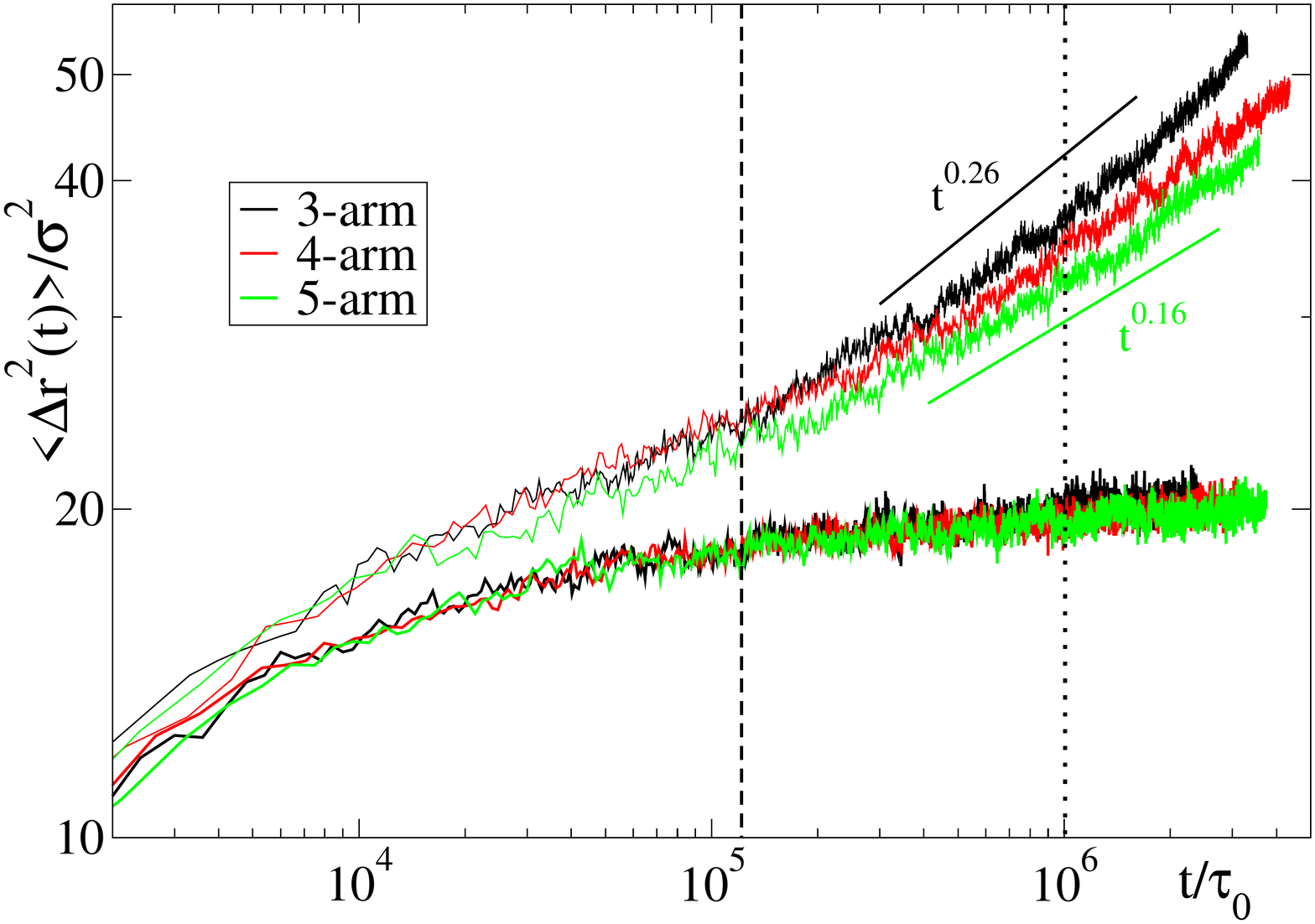}} \par}
\caption{Rescaled MSD of the centers of the star polymers with free (three top curves) and fixed ends (three bottom curves). See the main text for details of the rescaling procedure. The black and green solid lines indicate effective power-law behavior 
$\left< \Delta r^2 \right> \sim t^{0.26}$ and $t^{0.19}$, respectively.
The vertical dashed and dotted lines indicate the time scales equivalent 
to the limit of the NSE window ($t_{\rm NSE} = 400$ ns) 
and the limit of validity of DTD ($t^\ast = 3300$ ns, see subsection 4.2), respectively.}
\label{msd-scaled}
\end{figure}

The observed differences between the MSDs of the branch points are not only related to the different localization strength induced by the number of arms. Besides possible end-looping events that might be responsible for the marginal growth of the long-time plateau in the systems with fixed arm ends (\ref{msd}), ETD processes (e.g., diving modes) may lead to a broadening of the bare tube \cite{Bacova2013} that is also reflected in the MSD, and to a degree that may depend on the functionality. Since ETD is not related to constraint-release, broadening of the tube by ETD must stop at long times. Therefore irrespective of the specific $g(t)$ function describing the ETD-mediated broadening, $g(t)$ must reach an ultimate plateau. See Ref.~\cite{Bacova2013} for a detailed discussion on $g(t)$. With these ideas in mind, further insight in the relaxation of the branch point can be obtained by correcting for the former contributions through rescaling of the MSD.  Thus, for the stars with $f$ = 4 and 5 we rescale the MSD of the simulations with fixed ends to obtain the same plateau height as in the system $f$ = 3. We use the same scaling factors to correct the data of the respective simulations of $f$ = 4 and 5 with free ends. Moreover, to correct for  differences in the friction on the branch point, we scale the times so that all data sets coincide in the Rouse regime with the system $f$ = 3. \ref{msd-scaled} shows the corresponding data of the MSD after applying the rescaling procedure. In this representation all the data for the stars with fixed ends collapse, and the relaxation of the branch point in the systems with mobile ends is represented free of the $f$-dependent effects of localization and friction. The inspection of the data for the simulations with free ends reveals that the slope of the relaxation depends on the functionality. Thus, the data for $4 \times 10^5 \tau_0 < t < 2 \times 10^6 \tau_0$ can be effectively described by $\langle \Delta r^2\rangle \sim t^x$ with an exponent decreasing from $x \approx 0.26$ for $f = 3$ to $x \approx 0.19$ for $f = 5$. 
As it will be discussed later, this functionality dependence of the relaxation is not accounted for by current theories. \\



\begin{figure}[!ht]


{\centering \resizebox*{\textwidth}{1.65in}{\includegraphics{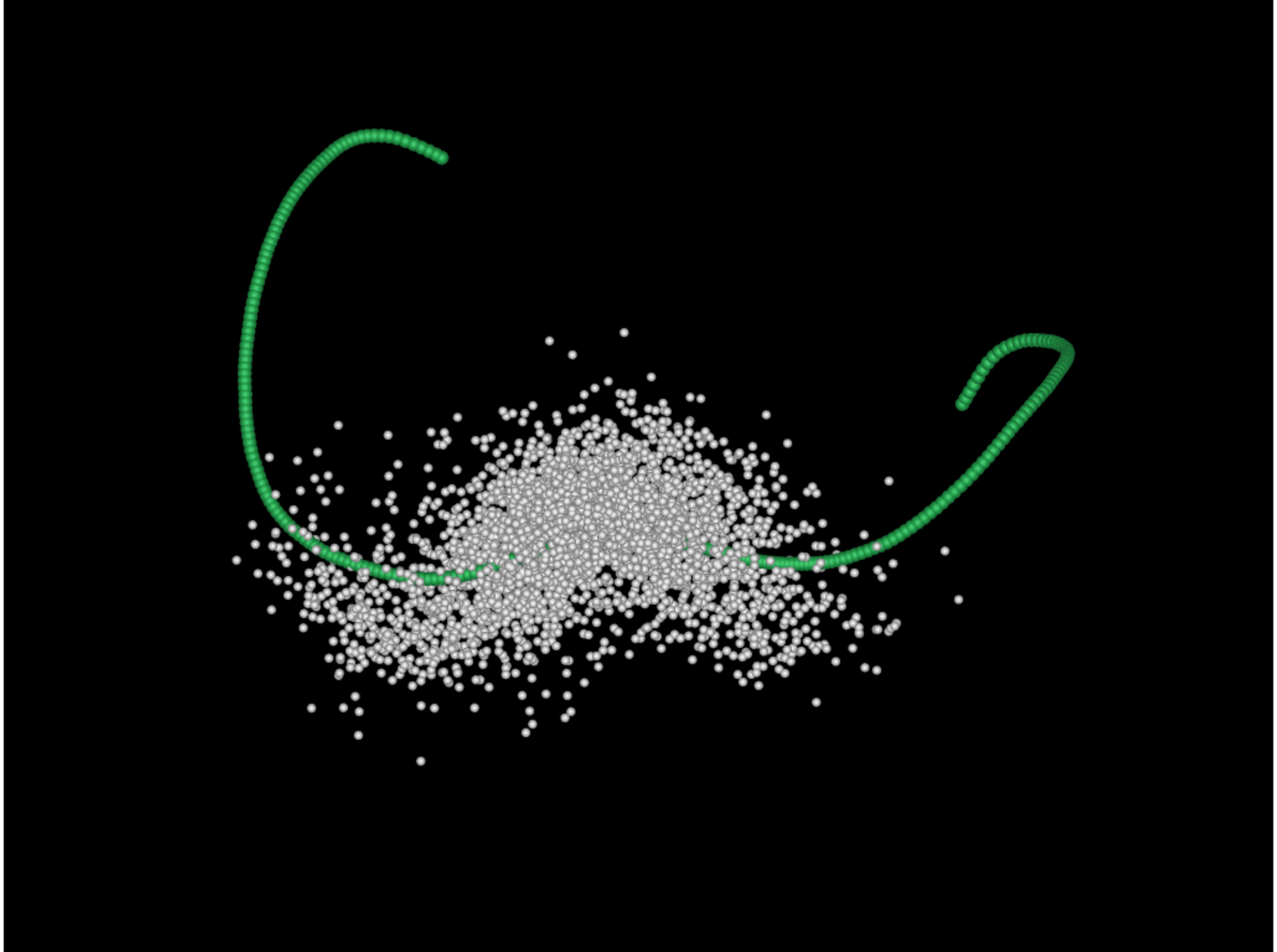}\includegraphics{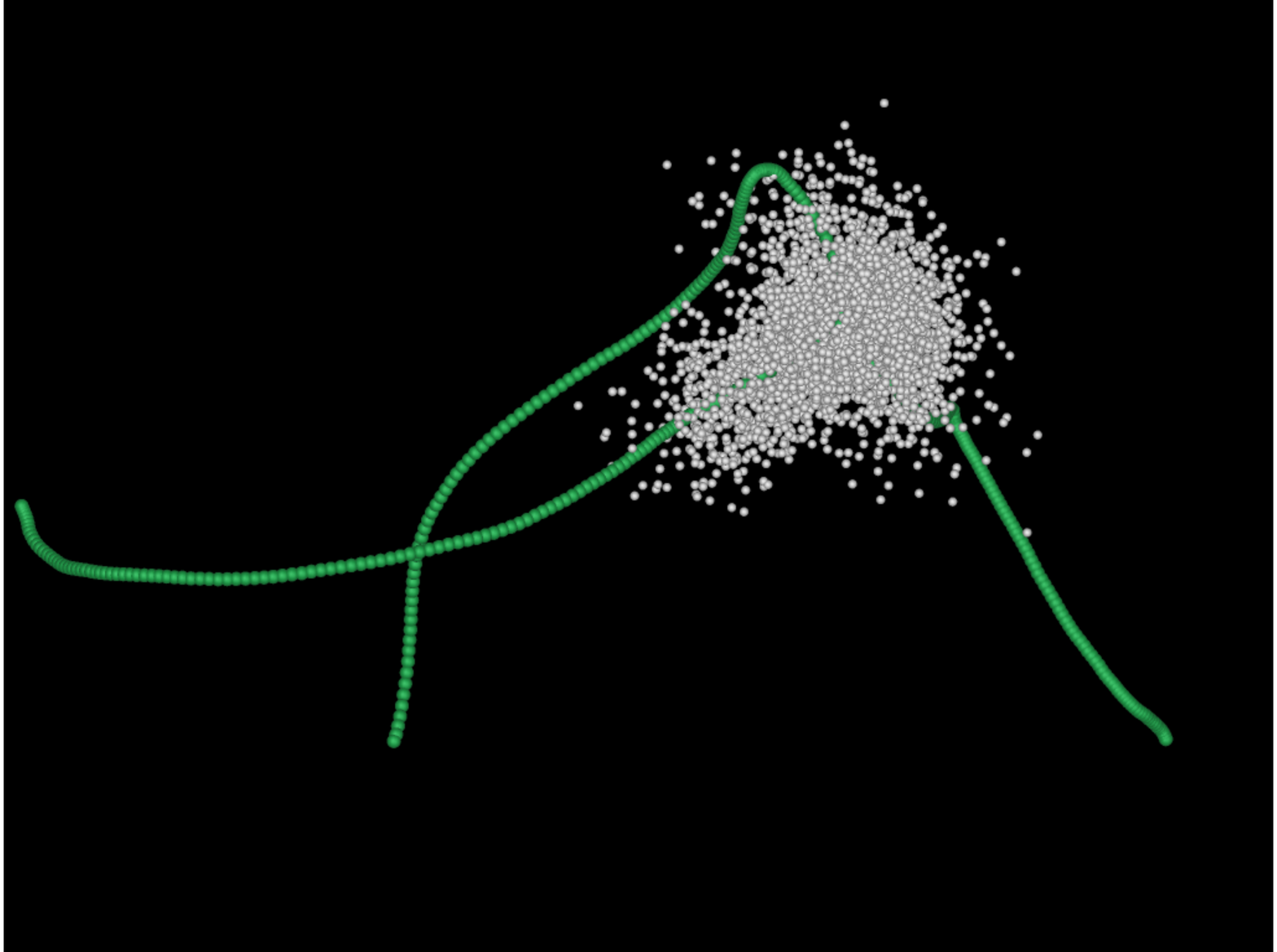}\includegraphics{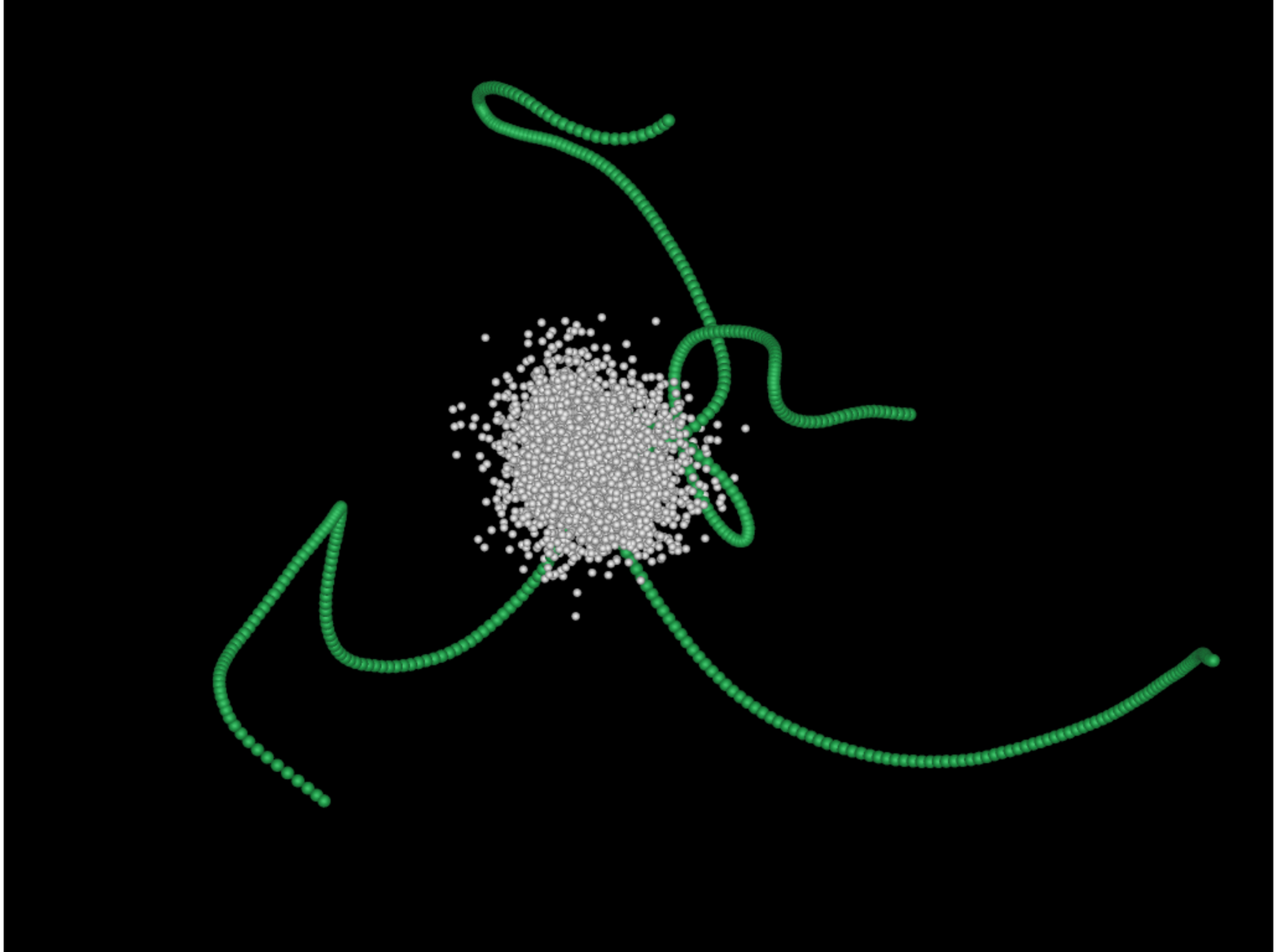}} \par}

\caption{Green lines: typical mean paths of a linear chain, a 3-arm-star and a 4-arm-star. The clouds of points represent the positions of the branch points at every time multiple of $\tau_{\rm e}/4$, in a trajectory of total time $t = 2 \times 10^6 \tau_0$.}
\label{cloud}
\end{figure}

Now we analyze the spatial confinement by characterizing the fluctuations of different segments of the arms in the simulations with fixed ends. We measure the fluctuations of each monomer around its position in the mean path. For this purpose we save,
for all the systems $2 \leq f \leq 5$ in the simulations with fixed ends, the coordinates of all the monomers at intervals of $t_{\rm save} = \tau_{\rm e}/4$. To compute the mean path we average for each monomer all its saved positions at $0 \leq t \leq t_{\rm max}$. For all the systems we use the same time $t_{\rm max} = 2 \times 10^6\tau_0$ (corresponding to the total time of the shortest run with fixed ends). By using the same values
of $t_{\rm save}$ and $t_{\rm max}$ we provide a fair comparison between the fluctuations around the mean path in the different systems.
The green curves in \ref{cloud} are typical mean paths of the arms of a linear chain, a 3-arm star, and a 4-arm star. The white points represent the positions in the real trajectory of the corresponding branch point. As can be seen, the increase of the functionality leads to a stronger degree of confinement of the branch point. Despite relaxation mechanisms being suppressed, in the linear chains ($f = 2$) with fixed ends the central monomer can still perform broad longitudinal motions along the tube path.  The motion of the branch point is already highly restricted in the case $f = 3$, though deeper explorations along the direction of the arms (diving modes \cite{Klein1986, Bacova2013}) can be found (see middle panel of \ref{cloud}). A stronger confinement of the branch points is observed for $f = 4$ and 5. We found no evidence of these deep diving modes in these systems. In principle, deep explorations of the tube involve some in-phase fluctuations of the arms around the branch point. These become very unlikely, due to the generated drag, already for relatively small functionalities, and for $f = 4$ and 5 they were not observed within the simulation time scale. 

\begin{figure}[!ht]
{\centering \resizebox*{0.5\textwidth}{2.5in}{\includegraphics{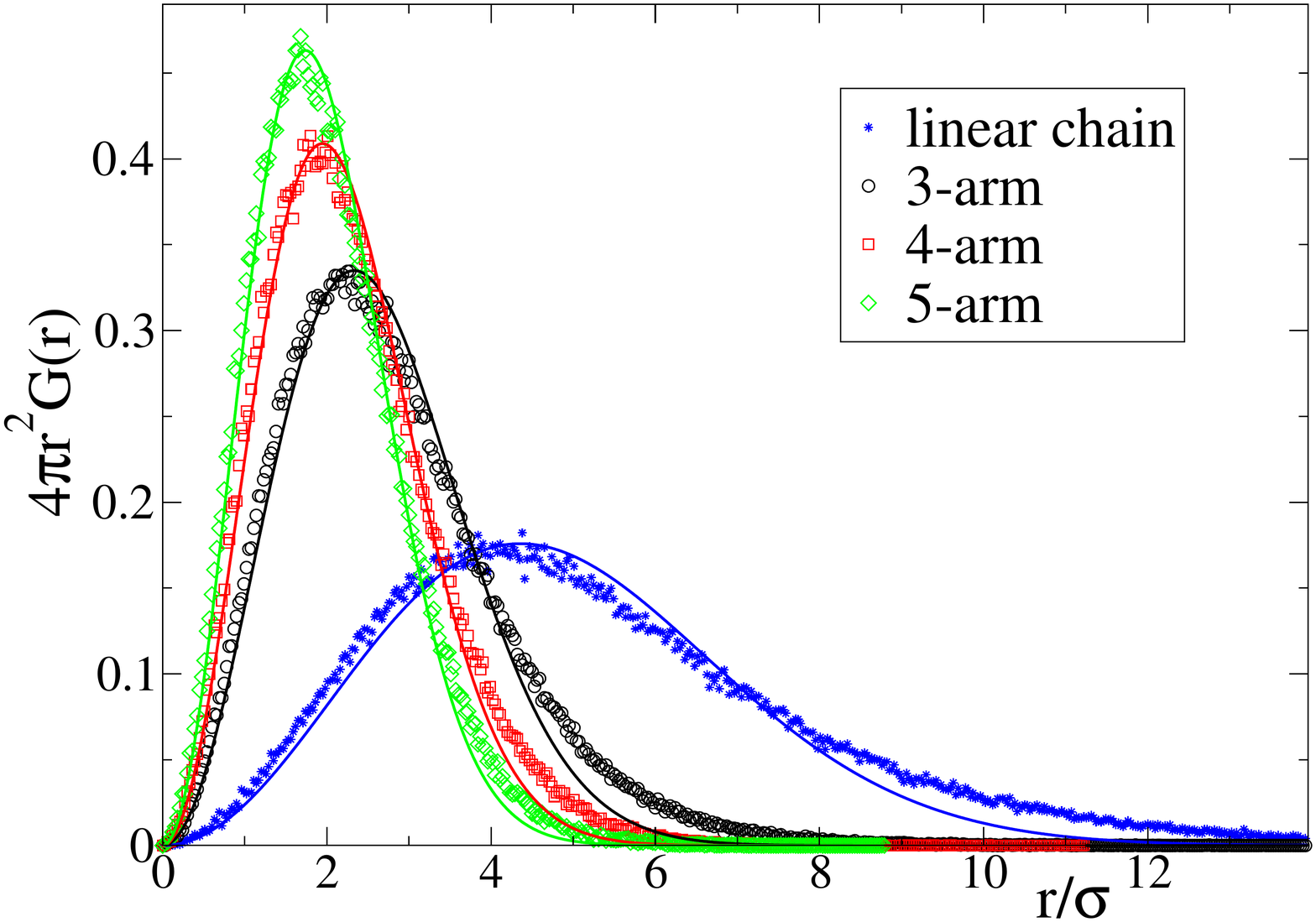}} \par}
\caption{Symbols: distribution of distances between the branch point and its position in the mean path. Lines: fits to a 3d-Gaussian function $G(r) = (3/2\pi \sigma_r^2)^{3/2}\exp[-3r^2/2\sigma_r^2]$.}
\label{dist-3-4-5-lin}
\end{figure}

\ref{dist-3-4-5-lin} shows the normalized distribution of distances $G(r)$ (multiplied by the phase factor $4\pi r^2$) from the branch point to its position in the mean path. As anticipated by the results in the previous figures, increasing the functionality leads to a stronger localization of the branch point, resulting in the narrowing and shifting of the distribution to smaller distances. 
In \ref{dist-3-4-5-lin} we show fits of the simulation data to 3d-Gaussian functions, $G(r) = (3/2\pi \sigma_r^2)^{3/2} \exp[-3r^2/2\sigma_r^2]$ with $\sigma_r$ the variance. Though a good description is achieved, the Gaussian function does not capture the significant tails that originate from the explorations of the tube, mainly for $f = 3$ (diving modes) and $f = 2$ (longitudinal motions). They are still present for higher $f$, but only weakly pronounced.

\begin{figure}[!ht]
{\centering \resizebox*{0.5\textwidth}{2.5in}{\includegraphics{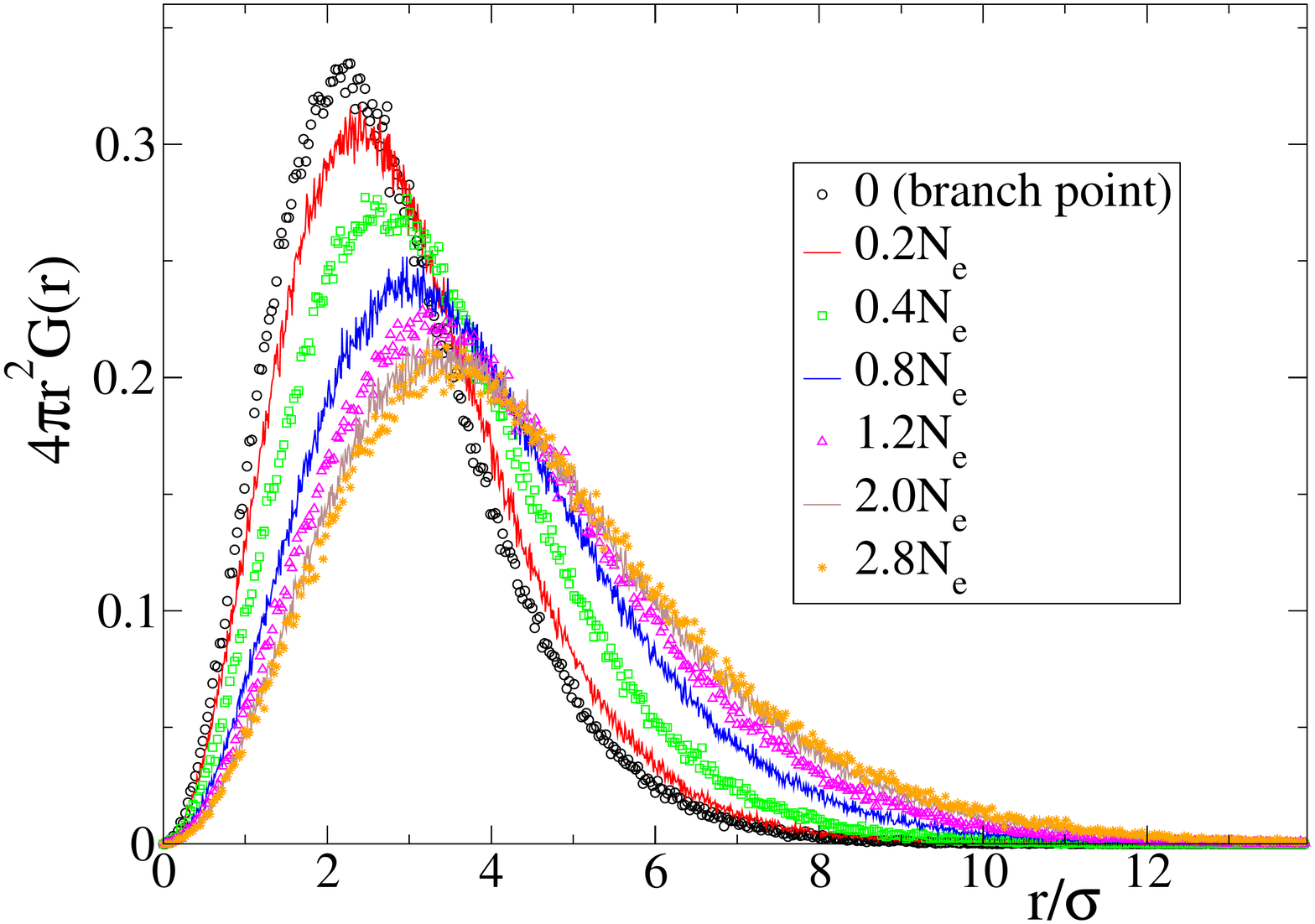}} \par}
\caption{Distributions of distances between selected monomers of the 3-arm star and their respective positions in the mean path. Each distribution is labeled (see legend) by the chain contour distance (number of monomers) between the selected monomer and the branch point  (in units of the entanglement length $N_{\rm e} = 25$).}
\label{dist-3-arm}
\end{figure}

\begin{figure}[!ht]
{\centering \resizebox*{0.5\textwidth}{2.5in}{\includegraphics{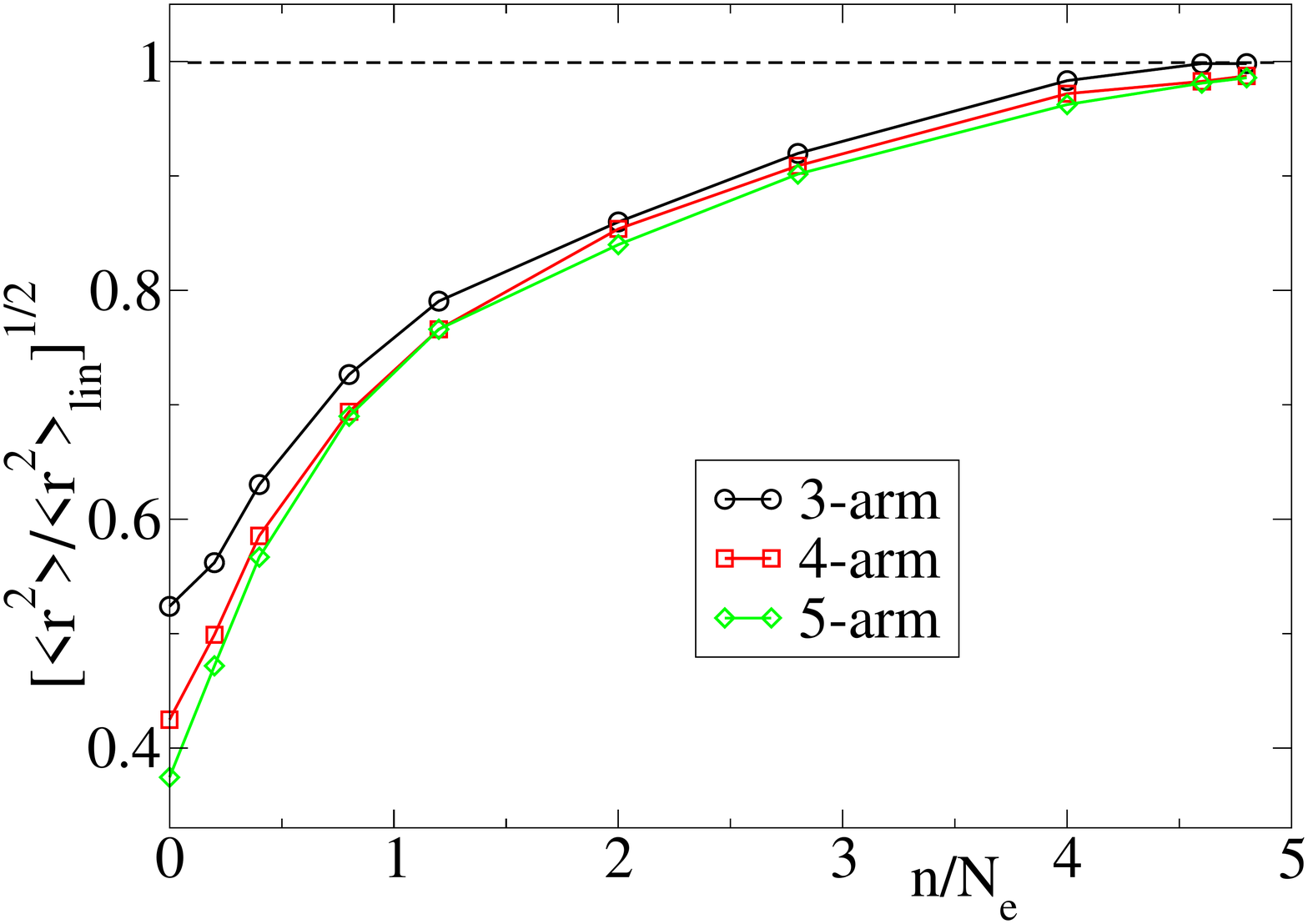}} \par}
\caption{Average fluctuation of selected monomers around their positions in the mean path, normalized by the corresponding value for the linear chains. The horizontal dashed line indicates the reference of the linear chain, $\left< r^2 \right> = \left< r^2_{\rm lin}\right>$. The data are represented vs the normalized contour distance, $n/N_{\rm e}$, to the branch point (so that $n/N_{\rm e} = 0$ and $n/N_{\rm e} = 5$ correspond to the branch point and the arm end, respectively).}
\label{fluctuations}
\end{figure}

\ref{dist-3-arm} shows the corresponding distributions $4\pi r^2 G(r)$ for several selected monomers in the 3-arm stars. The distributions are labeled by the chain contour distance (number of monomers, $n$) between the selected monomer and the branch point, measured in units of the entanglement length $N_{\rm e}$. The distributions become broader and are shifted to larger distances, reflecting a weaker confinement, by moving away from the branch point. At distances larger than two entanglements from the branch point the distributions become very similar to each other, suggesting that the influence of the branch point has a reach of roughly 2-3 entanglements. 
This is confirmed in \ref{fluctuations}, which shows for several selected monomers and all the investigated stars, the average fluctuation $\langle r^2 \rangle^{1/2}$ around the mean path, 
with $\langle r^2\rangle = \int_0^\infty 4 \pi r^2 G(r)dr$. The data have been normalized by the respective values of the linear chains, and are represented vs the normalized contour distance $n/N_{\rm e}$ to the branch point, so that $n/N_{\rm e} = 0$ and $n/N_{\rm e} = 5$ correspond to the branch point and the arm end, respectively. By looking at the data of \ref{fluctuations}, in the neighborhood of the branch point ($n/N_{\rm e} \rightarrow 0$) we find that the stronger confinement of the monomers in the stars with respect to the linear case (dashed line in \ref{fluctuations}) is not limited to the innermost segments. This feature is also observed even at distances of 2-3 entanglements from the branch point.

\subsection{Tube relaxation}

\begin{figure}[!ht]
{\centering \resizebox*{0.5\textwidth}{2.5in}{\includegraphics{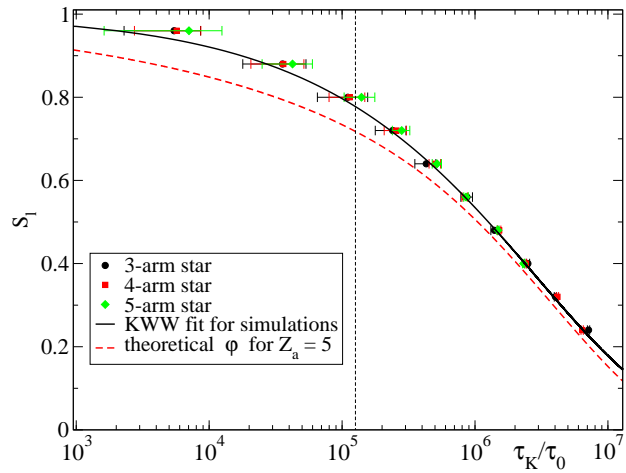}} \par}
\caption{Symbols: Unrelaxed path coordinate $s_l$ (decreasing from 1 at the arm end to 0 at the branch point) vs the KWW relaxation time of the tangent correlator $\varphi_l$. The black solid curve is a KWW fit. The dashed red line is the theoretical $\varphi$ for $Z_{\rm a}=5$,
obtained by using \ref{eq:tau-early,eq:u-eff,eq:tau-late,eq:taux,eq:tube-surv}. 
The vertical dashed line indicates the time scale equivalent to the limit of the NSE window, $t_{\rm NSE} = 400$ ns (see below).}
\label{tube-surv}
\end{figure}

Now we compare the relaxation of the tube in the investigated stars by analyzing their tube survival probability functions. Following the original work of Doi and Edwards \cite{Doi1988}, the tube survival probability can be formulated in terms of the tangent correlation function, $\varphi_l(t)$. Following the arguments of Ref.~\cite{Bacova2013} this was computed for each arm, in the simulations with free ends, through the expression: 
%
%
%
\begin{equation}
\varphi_l(t) = \left< \vec{u}_{i,l}(0)\cdot\left( \vec{R}_i^{\rm e}(t) - \frac{1}{f-1} \sum_{i\neq j}^{f} \vec{R}_j^{\rm e}(t) \right) \right>
\end{equation}
Here $\vec{u}_{i,l}(t)$ is the tangent vector of the $l$th segment of the $i$th arm at time $t$. 
$\vec{R}_i^{\rm e} (t)$ is the end-to-end vector of the $i$th arm. The sum within the parenthesis includes all the other arms $j \neq i$ in the same star, each arm carrying a cross-correlation factor $1/(f-1)$. The brackets denote average over the $f$ arms of the same star (since they are identical), and over all the stars in the system. The arms were divided in 12 segments of 10 monomers (labelled $l$ = 1, 2,..., $n_{\rm s}$ = 12 from the branch point to the arm end), and the end-to-end vectors of such segments were used as the tangent vectors $\vec{u}_{i,l} (t)$. 
We define the discretized tube coordinate as $s_l = l/n_{\rm s}$ , 
so that it changes from $s_l \approx 0$ to $s_l =1$ by moving from the branch point to the arm end. The tangent correlators were fitted to stretched exponential (Kohlraus-William-Watts, KWW) functions $ \varphi_l (t)=\exp(-(t/\tau_{\rm K} )^\beta )$. In \ref{tube-surv} we represent  $s_l$ vs. the KWW relaxation time $\tau_{\rm K}$ of the respective tangent correlator $\varphi_l$. The three stars show the same long-time relaxation of the tube coordinate. This is clearly nonexponential and can be described by the same KWW function $\varphi_l (t)$ (black solid curve), with $\beta = 0.44$ and $\tau_{\rm K} = 2.9\times 10^6 \tau_0$, for the three investigated stars. This KWW function can be used to model the relaxation of the continuous tube coordinate $s$. Since all the arms of all the stars in the system are identical, the tube survival probability $\varphi(t)$, which accounts for the total fraction of unrelaxed material, 
is identical to the $s(t)$ represented in \ref{tube-surv} and is described by the former KWW function.

Now we discuss the consequences of these observations for DTD. First, it must be noted that 
there is a limitation in the time scale for the validity of DTD \cite{McLeish2002}. Thus, DTD can only be valid at times $t$ for which the rate for tube broadening is slower than the rate for self-diffusion of the monomers on the segments relaxing at $t$ (otherwise chain relaxation would not be impeded by the tube). 
This criterion is quantitatively given by the expression \cite{McLeish2002}:
\begin{equation}
(1-s)s^2 > \frac{2}{3Z_{\rm a}}
\label{eq:dtdvalid}
\end{equation}
For the simulated stars $Z_{\rm a}=5$, so that DTD can only be invoked for $s > 0.55$. This range of tube coordinate corresponds to the time window 
$t < t^\ast = 10^6 \tau_0$ in the tube survival probability displayed in \ref{tube-surv}. Though the total simulation time is longer than $t^\ast$, the differences in the relaxation of the branch point are already clearly visible at $t < t^{\ast}$  (see \ref{msd-scaled}).
According to DTD theory, the tube parameters are renormalized by factors that depend only on the tube survival probability (see below). Therefore if the theory is correct DTD, in its time scale of validity $t < t^{\ast}$, should affect the relaxation of the three investigated stars identically, since they show the same $\varphi(t)$ (\ref{tube-surv}). As discussed in Ref.~\cite{Bacova2013}, further renormalization is necessary in order to include non-constraint release ETD (diving into the adjacent arm tubes).  
The results for the MSD in \ref{msd-scaled} have already been corrected for ETD and differences in the friction and the localization strength. Therefore, the different slopes in the data of \ref{msd-scaled} are the signature of a different effect of DTD in the relaxation of symmetric stars with identical arms but different functionality, in spite of their identical DTD renormalization.  This observation is in contradiction with the theoretical expectation.

The simulation results for the tube survival probability can be compared with the theory introduced by Milner and McLeish (see e.g., Refs.~\cite{Milner1998,Inkson2006} for details and analytical expressions), which considers arm retraction and DTD. 
The two time scales that have to be considered for the calculation of the tube survival probability are the early arm relaxation time $\tau_{\rm early}$ (at $t<\tau_{\rm e}$) and the activated retraction time (at $t>\tau_{\rm e}$). The $\tau_{\rm early}$ is expressed as
%
\begin{equation}
\tau_{\rm early}(s)=\frac{225\pi^3}{256} \tau_{\rm e} (1-s)^4 Z_{\rm a}^4 
\label{eq:tau-early} 
\end{equation}
%
%
%
with $s$ the arm tube coordinate (decreasing from $s =1$ at the arm end to 0 at branch point). The activated retraction time,
or `late' relaxation time $\tau_{\rm late}(s)$,
is the time needed for an arm to retract a distance $1-s$ in its effective potential $U_{\rm eff}$. This is given by  

\begin{equation}
U_{\rm eff}(s)=\frac{15 Z_{\rm a}}{4} \frac{1-s^{1+\alpha}\left[ 1+\left(1+\alpha\right)(1-s)\right]}{\left(1+\alpha \right) \left(2+\alpha \right)}
%
%
\label{eq:u-eff}
\end{equation}

\noindent
The scaling exponent $\alpha$ can be chosen as 1 or 4/3 (see discussion in Ref.~\cite{Milner1997}). The late relaxation time is obtained as \cite{Milner1998}:

\begin{equation}
\tau_{\rm late}(s) \approx \tau_{\rm e} Z_{\rm a}^{3/2} \left(\frac{\pi^5}{30}\right)^{1/2} \times 
\frac{\exp\left[ U_{\rm eff}(s)\right]}
{(1-s)\left[ s^{2\alpha} + \left(  \frac{4 Z_{\rm a}}{15}   \left( 1+ \alpha \right) \right)^{2\alpha/\left(\alpha+1 \right)} \Gamma \left( \frac{1}{\alpha+1} \right)^{-2} \right]^{1/2}}
%
%
\label{eq:tau-late}
\end{equation}
with $\Gamma$ the Euler function.
Finally we can calculate the complete relaxation time scales of the arms with an expression that describes the crossover from early Rouse-like relaxation (time scale $\tau_{\rm early}(s_x)$) 
to late activated relaxation (time scale $\tau_{\rm late}(s_x)$).

%
\begin{equation}
\tau(s) = \frac{\tau_{\rm early}(s) e^{U_{\rm eff}(s)}}{1+\tau_{\rm early}(s) e^{U_{\rm eff}(s)}/\tau_{\rm late}(s)}
%
\label{eq:taux}
\end{equation}
\noindent 
The local survival probability \cite{McLeish2002} at any position of the arm is described by $\delta(t,s) \sim \exp(-t/\tau(s) )$. Integrating  $\delta(t,s)$ over the complete arm length we can calculate the tube survival probability:
\begin{equation}
\varphi(t) = \int_{0}^{1} \exp(-t/\tau(s)) ds
\label{eq:tube-surv} 
\end{equation} 
The theoretical result for stars with arm length $Z_{\rm a} =5$ is shown in \ref{tube-surv} (red dashed curve). The theory provides a reasonable description of the simulation results at times longer than the Rouse time of the arm, $\tau_{\rm R} = \tau_{\rm e}Z^2_{\rm a} \approx 5 \times 10^4 \tau_0$.

\section{NSE results}

In order to investigate the effect of functionality and the arm length on the mobility of the branch point we performed NSE experiments with 3- and 4-arm symmetric polyethylene stars. We measured the dynamic structure factor for the small ($Z_{\rm a} = 5$) and large ($Z_{\rm a} = 13$)  stars. Different $q$ values between $q=0.05\text{\AA}^{-1}$ and $0.115\text{\AA}^{-1}$ were measured for times up to around 400 ns. As aforementioned, the protonated label was the same in all 
the samples, providing a correct comparison between them.

\begin{figure}[!ht]
{\centering \resizebox*{\textwidth}{2.5in}{\includegraphics{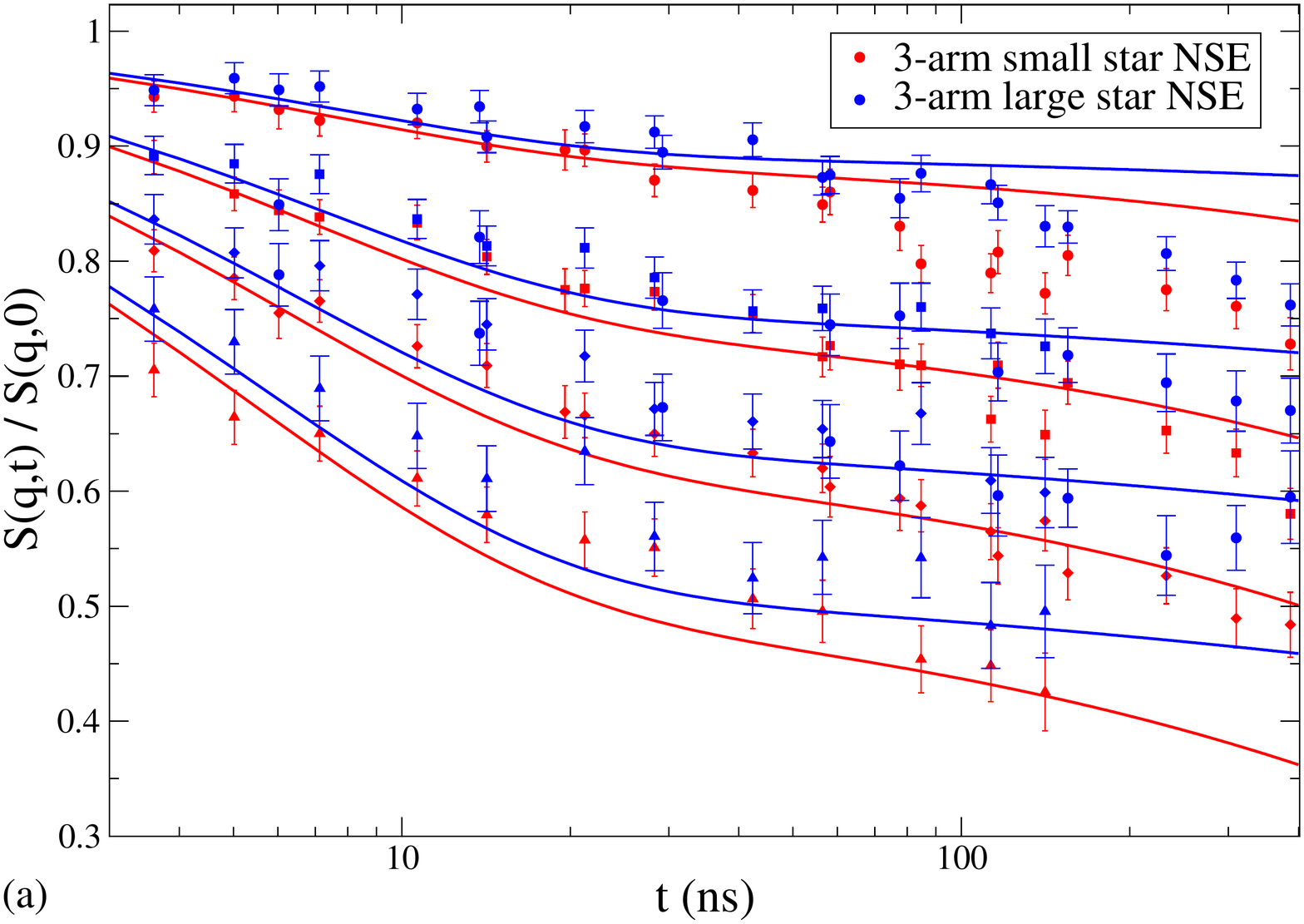}\includegraphics{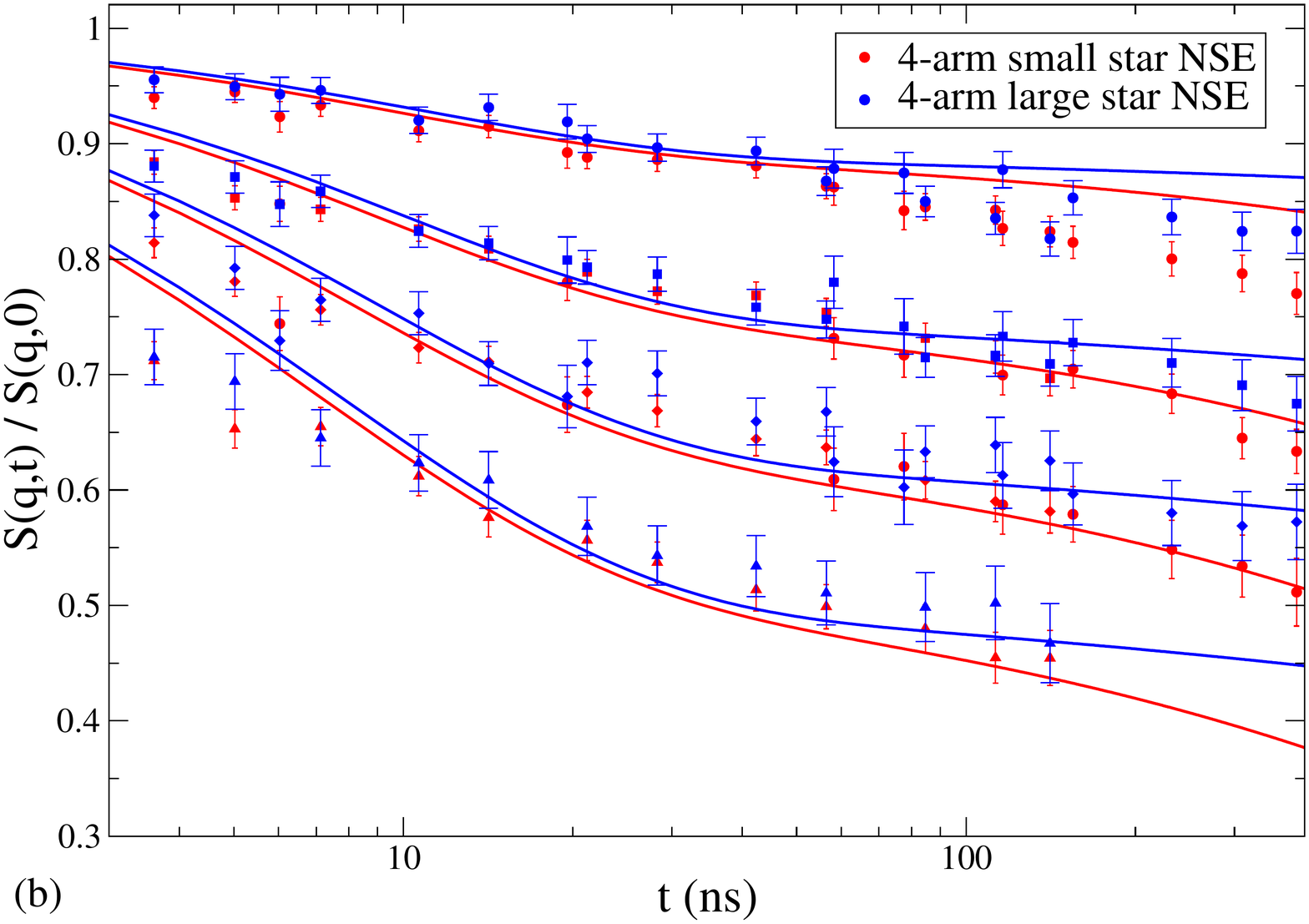}} \par}
\caption{Measured dynamic structure factor of large $Z_{\rm a} =13$ (blue) and small $Z_{\rm a}=5$ (red) stars. (a): 3-arm stars. (b): 4-arm stars. Symbols are NSE data, lines are fits to the Vilgis-Bou\'{e} theory with DTD. The $q$ values are $0.05 \text{\AA}^{-1}$, $0.077 \text{\AA}^{-1}$, $0.096 \text{\AA}^{-1}$ 
and $0.115 \text{\AA}^{-1}$ from top to bottom.}
\label{dyn-short-long}
\end{figure}

The normalized  dynamic structure factors, $S(q,t)/S(q)$, with $S(q) = S(q,0)$, are shown in \ref{dyn-short-long}. After the Rouse-like decay up to the entanglement time, a plateau arises, followed by a decay at long times.
By comparing the smaller, moderately entangled 3-arm stars to their longer counterparts, one can easily see that $S(q,t)/S(q)$ exhibits a faster decay in the small stars, reflecting a higher mobility of the branch point (see \ref{dyn-short-long}a). Up to around 30 ns both stars show very similar behavior and the difference becomes clear only at longer times. The same behavior is observed for the 4-arm stars (\ref{dyn-short-long}b), with the same result of a stronger localization of the branch point in the large stars. 
The arm relaxation times are far beyond the NSE time window both for the short and long arm stars (see below).
Therefore, we tentatively assign the observed decays at the long NSE times to broadening of the tube, this being stronger
in the short stars than in their large counterparts with the same functionality. 



\begin{figure}[!ht]
{\centering \resizebox*{\textwidth}{2.5in}{\includegraphics{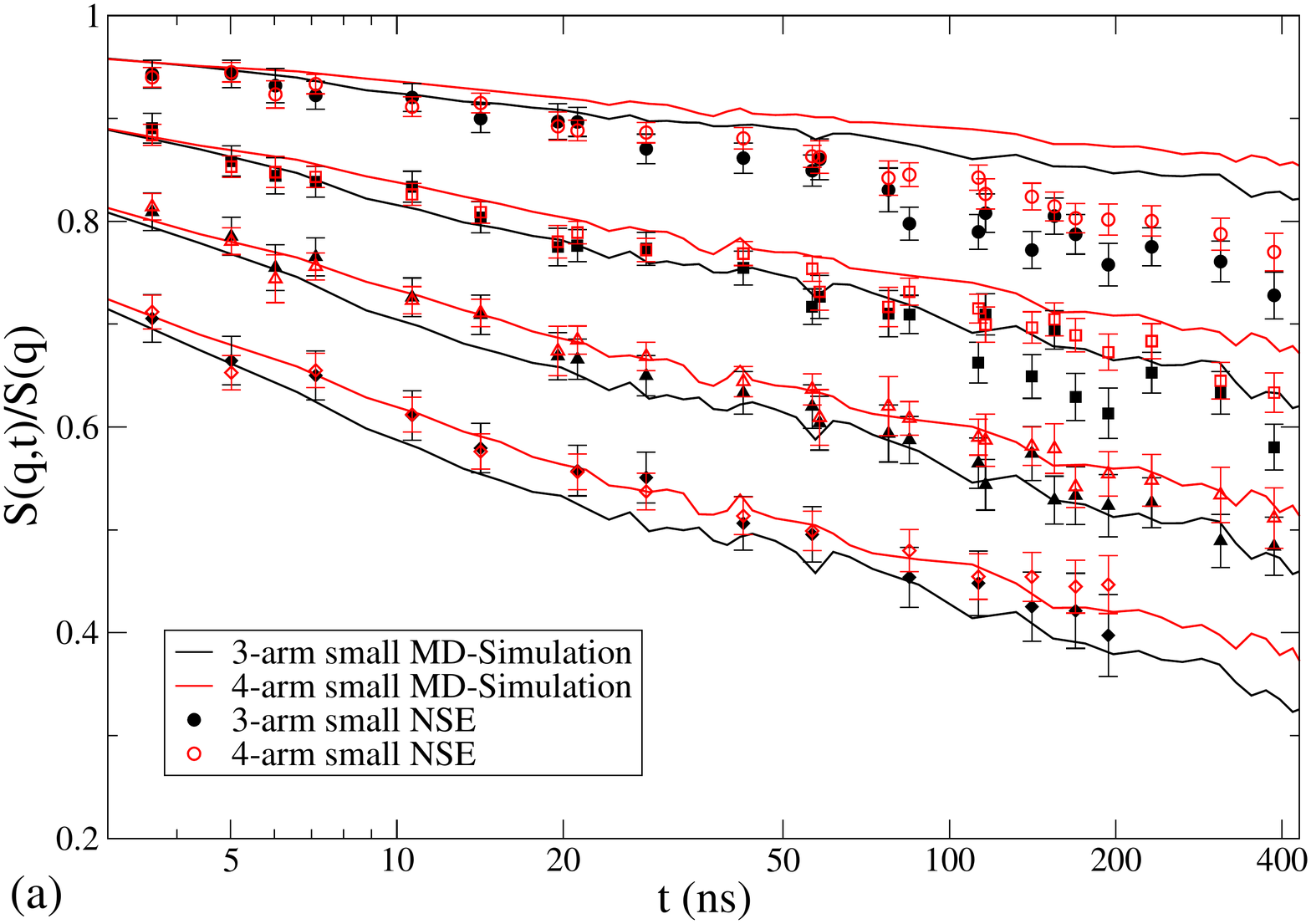}\includegraphics{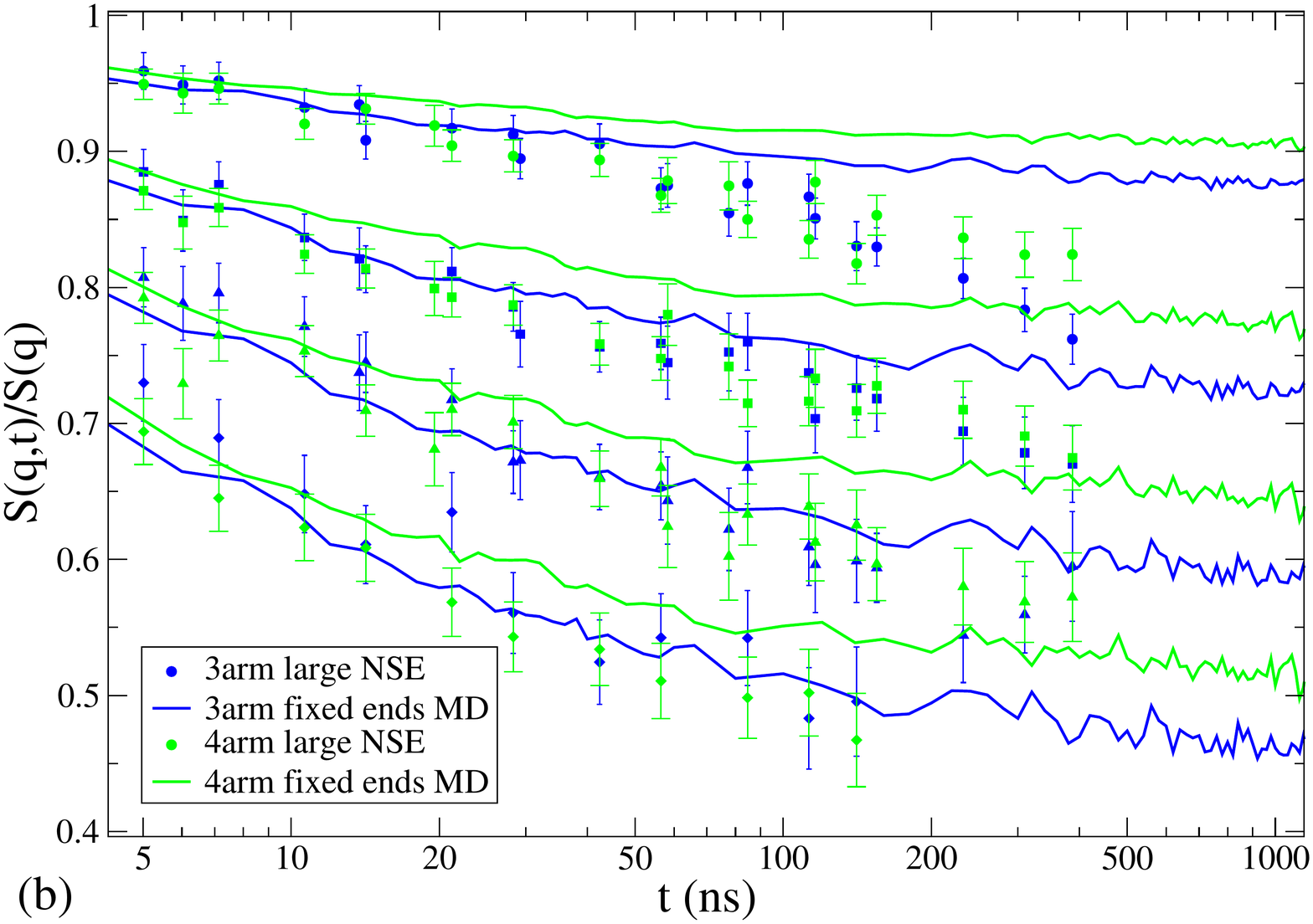}} \par}
\caption{(a): Dynamic structure factor from NSE (symbols) and MD simulations with rescaled units (lines), for small 3-arm (black) and small 4-arm (red) stars. (b): Dynamic structure factor from NSE (symbols) and MD simulations with rescaled units (lines) for large 3-arm (blue) and large 4-arm (green) stars. In both panels the $q$ values are (in $\text{\AA}^{-1}$): 0.05 (circles), 0.077 (squares), 0.096 (triangles) and 0.115 (diamonds), from top to bottom.}
\label{dyn-MD-NSE}
\end{figure}

\begin{figure}[!ht]
{\centering \resizebox*{0.5\textwidth}{2.5in}{{\includegraphics{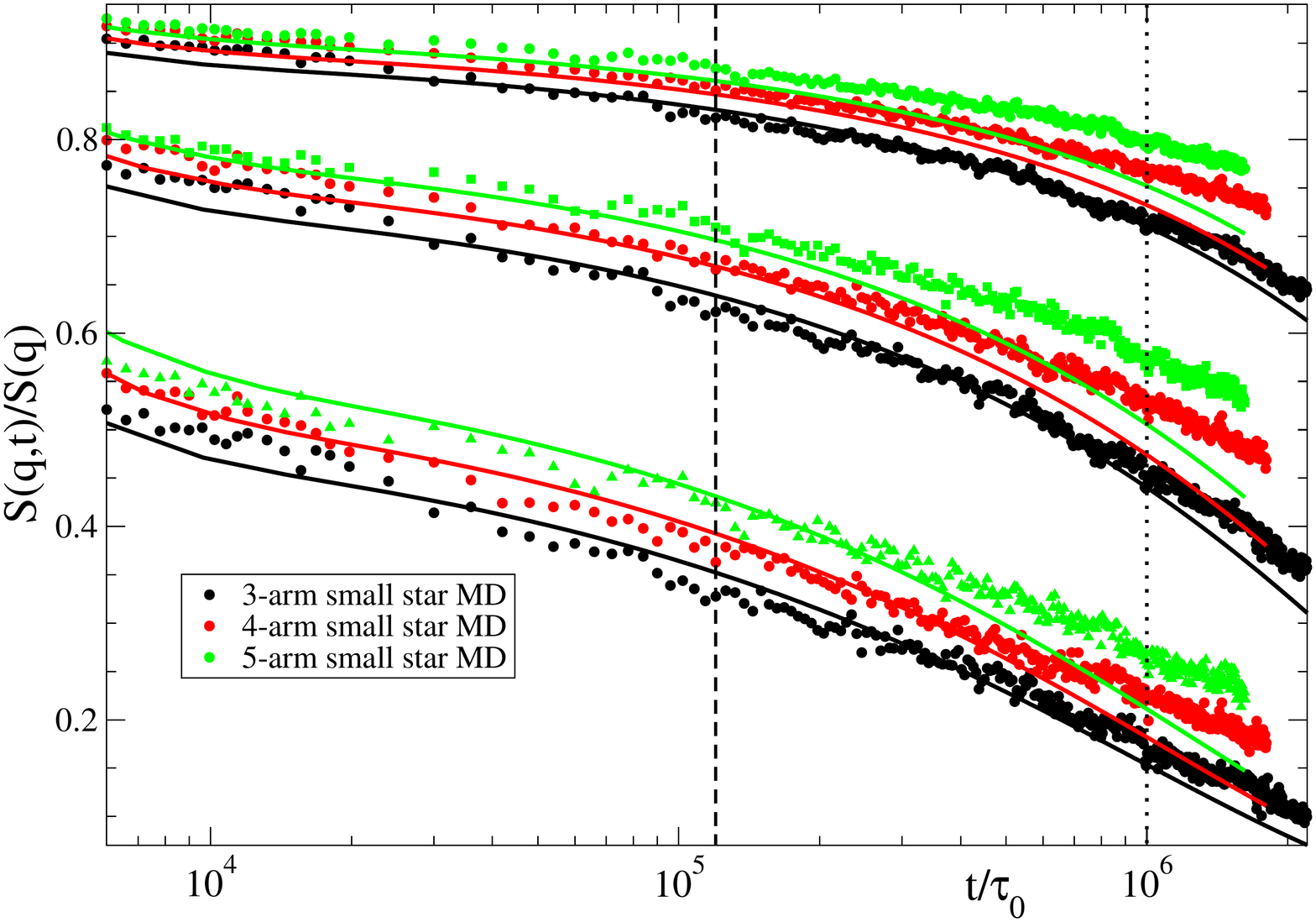}}} \par}
\caption{Dynamic structure factor of small 3-arm (black), 4-arm (red) and 5-arm (green) stars. Symbols are the MD data. 
Lines are fits to the theory (Vilgis-Bou\'{e} model with DTD), with $q\sigma =$ 0.28 (circles), 0.43 (squares) and 0.54 (triangles), from top to bottom. 
These wave vectors correspond to the NSE values (in $\text{\AA}^{-1}$) $q = 0.05$, 0.077 and 0.096, respectively. 
The vertical dashed an dotted lines indicate the time scales equivalent 
to the limit of the NSE window ($t_{\rm NSE} = 400$ ns) 
and the limit of validity of DTD ($t^\ast = 3300$ ns), respectively.}
\label{dyn-simu-theo}
\end{figure}

A direct comparison between the NSE spectroscopy results and the MD simulations is shown in \ref{dyn-MD-NSE}a. The coherent dynamic structure factor for the simulated systems has been calculated through \ref{eq:s-q-t}.
Namely we only include in the calculation the innermost monomers
of three arms, with a size of half an entanglement per arm (12 beads), i.e., mimicking the labeling of the experiments.
The simulation results have been scaled to the experimental times (in ns) and wave vectors (in $\text{\AA}^{-1}$) by factors corresponding to the ratios of the corresponding entanglement times and tube diameters. Therefore the simulation time has been rescaled by the factor  $\gamma_t = 7/2100$ ns,  where 7 ns and $2100 \tau_0$ are approximate values of the experimental \cite{Zamponi2010} and simulation \cite{Zhou2007,Bacova2013} entanglement time, respectively. We adjusted the entanglement time slightly to optimize the match of the MD and NSE data.
The simulation $q$ values have been rescaled by the factor $\gamma_q = 8.75/49$ $\text{\AA}^{-1}$, with 49 $\text{\AA}$ and $8.75\sigma$ approximate values of the tube diameter of the linear polyethylene \cite{Wischnewski2002} and the simulated bead-spring chains \cite{Everaers2004}, respectively.
With the used scaling factors for the time and wave vector we achieve a good agreement between MSD and NSE for the small stars (see \ref{dyn-MD-NSE}a),
except for $q=0.05 \text{\AA}^{-1}$, where the experimental results show a much stronger drop in $S(q,t)/S(q)$ than the simulations. A similar observation at $q=0.05 \text{\AA}^{-1}$ is also found between the NSE data and the Vilgis-Bou\'{e} theory with DTD 
(see \ref{dyn-short-long} and discussion below). The reason for such discrepancies observed only at  $q=0.05 \text{\AA}^{-1}$ is not clear. It is worth mentioning that, in the case of the simulations, there is a good agreement with the theory in all the equivalent NSE range of time and wave vectors (see \ref{dyn-simu-theo} and discussion below). Therefore we tend to believe that the mentioned discrepancies 
in the NSE data at $q=0.05 \text{\AA}^{-1}$ might be just an artifact of the experimental setup (as, e.g., non-negligible contributions from incoherent scattering).

Though, due to their high computational cost, we have not simulated the counterparts of the experimental long stars ($Z_{\rm a}=13$), 
it is interesting to compare the NSE data of the latter with the simulations
of the small stars with fixed ends (see \ref{dyn-MD-NSE}b). As expected, suppressing relaxation by fixing the arm ends leads to a plateau in the $S(q,t)$.
A clear decay from the plateau  is instead found in the experimental long stars, reflecting the effect of progressive delocalization through DTD,
which is absent in the case of the stars with fixed arm ends.

In order to describe the arm length dependence of DTD we modify the model of Vilgis and Bou\'{e} \cite{Vilgis1988}. This model was initially designed for cross-linked networks and describes the dynamics of confined chain segments in a harmonic potential. The model was previously applied to describe the confinement for branch points in large star polymers \cite{Zamponi2010}. The relative mean square displacement of two monomers $m$ and $n$ is calculated as:
\begin{align}
\left< \left( \vec{r}(m,t) - \vec{r}(n,0) \right)^2 \right> = 
3R_{\rm mesh}^2
\bigg[ 
1-\frac{1}{2}
\bigg(
	2 \cosh
	\left( \frac{l^2}{3R_{\rm mesh}^2} |n-m| 
	\right) \nonumber \\
	-\exp
	\left( 
		-|n-m| \frac{l^2}{3R_{\rm mesh}^2} 
	\right) 
	{\rm erf}
	\left( 
		\frac{1}{3R_{\rm mesh}^2}\sqrt{Wl^4t} - |n-m| \frac{l^2}{2\sqrt{Wl^4t}}
	\right) \nonumber \\
	-\exp
	\left( 
		|n-m| \frac{l^2}{3R_{\rm mesh}^2} 
	\right) 
	{\rm erf}
	\left( 
		\frac{1}{3R_{\rm mesh}^2}\sqrt{Wl^4t} + |n-m| \frac{l^2}{2\sqrt{Wl^4t}} 
	\right)
\bigg)
\bigg]
\label{eq:vilgis}
\end{align}
with $\rm erf()$ the error function, $l$ the segment length, $Wl^4$ the Rouse rate, and $R_{\rm mesh}$ the well parameter, which corresponds to the radius of gyration of the single mesh. By using the MSD from \ref{eq:vilgis}, the theoretical dynamic structure factor can be calculated in Gaussian approximation:
\begin{equation}
S(q,t) = \sum_{m,n}\exp(-q^2 \langle ( \vec{r}(m,t) - \vec{r}(n,0) )^2 \rangle/6)
\label{eq:gaussian}
\end{equation}

The Vilgis-Bou\'{e} model only accounts for fluctuations of the branch points and does not include relaxation.
Therefore, the dynamic structure factor obtained from \ref{eq:vilgis} and \ref{eq:gaussian} goes into a plateau which signifies the confinement of the branch point. In order to describe the gradual loss of confinement by DTD we introduce a time dependent broadening
of the mesh size. Namely, we have to adapt \ref{eq:vilgis} for the effect of DTD and other relaxation processes that are not accounted for in the theory. The tube parameters change over time with the following DTD renormalizations \cite{Milner1997,Milner1998,McLeish2002}:
\begin{equation}
d(t) =  d\varphi^{-\alpha/2}(t) 
\label{eq:renormtube}
\end{equation}
\begin{equation}
|n-m|(t) =  |n-m|\varphi^{\alpha}(t)
\label{eq:renormdis}
\end{equation}
where $\varphi(t)$ is the dilution function and  $\alpha = 1$ or 4/3 is the scaling exponent. The mesh size in the Vilgis-Bou\'{e} model is related to confining tube through \cite{Zamponi2010} $R^2_{\rm mesh} = d^2 /6$.
Ideally $\varphi(t)$ is the tube survival probability. However, in order to account for the ETD not included in the theory we rescale by an additional time-dependent function \cite{Bacova2013}:
$\varphi^{\alpha}(t) \rightarrow g(t)\varphi^{\alpha}(t)$. We use the ETD function $g(t)$ obtained from the simulations of 3-arm stars 
with fixed ends (see Ref.~\cite{Bacova2013}), applying the time scaling factor $\gamma_t$ to transform to experimental units. 
In the case of $\varphi(t)$, both for the simulated and experimental small stars ($Z_{\rm a}=5$)
we use the KWW fitting function obtained from the simulations (black curve in \ref{tube-surv}).
Therefore we used the stretching exponent 
$\beta = 0.44$, and the KWW time $\tau_{\rm K} = 2.9\times 10^6 \tau_0$ for the simulations and (after rescaling by $\gamma_t$) 
$\tau_{\rm K} = 9700$ ns for the NSE data.
In the case of the long stars ($Z_{\rm a}=13$) we used  \ref{eq:tau-early,eq:u-eff,eq:tau-late,eq:taux,eq:tube-surv} to construct $\varphi(t)$. The resulting function was fitted to a KWW function with a stretching exponent $\beta=0.265$ and a KWW time $\tau_{\rm K} = 3.7 \times 10^8\tau_0$ 
(corresponding to 1.2 $\times 10^6$ ns in the NSE units).
The obtained relaxation times $\tau_{\rm K}$ for the tube survival probability, which provide an estimation of the arm relaxation times, are far beyond the NSE time window ($t_{\rm NSE} = 400$ ns). Moreover, the NSE window is in both cases within the time limit $t < t^\ast$ of validity of DTD.
As aforementioned, for $Z_{\rm a} = 5$ this time is $t^\ast = 10^6 \tau_0$, corresponding to 3300ns. In the case of $Z_{\rm a}=13$, according to \ref{eq:dtdvalid} DTD is valid for $s > 0.25$. This corresponds to $t < t^\ast = 10^9 \tau_0$ and $3.3\times 10^6$ns in MD and NSE units, respectively.
In summary, the values of the former time scales justify assigning the observed decays at $\tau_{\rm e} < t < t_{\rm NSE}$ to the dilation of the tube probed by the branch point.

We apply the DTD renormalization given by \ref{eq:renormtube,eq:renormdis}, with the simulation and theoretical inputs as explained above, to the MSD of \ref{eq:vilgis}. 
Finally, we insert the renormalized MSD in \ref{eq:gaussian} to compute the theoretical dynamic structure factor via an intra-arm sum over the protonated label, and an average over all the arms. The value of the Rouse rate $Wl^4$ for each system can be obtained from
the known value for the linear chains \cite{Richter1993} $Wl^4 = 7 \times 10^4 \text{\AA}^4 {\rm ns}^{-1}$. This is rescaled by the corresponding
factor $2/f$ to get the Rouse rate of the stars \cite{Zamponi2010}. 
It is worth mentioning that the renormalization of the intramolecular distance $|n-m|$ has a very small effect, in comparison with
the renormalization of the tube diameter, in the resulting theoretical dynamic structure factor. Likewise, including the ETD function
$g(t)$ in the renormalization of the tube, or just taking $g(t)=1$, has a negligible effect in the theoretical $S(q,t)$ of the 4- and 5-arm stars.
The results presented here correspond to calculations with scaling exponent $\alpha =1$ in \ref{eq:u-eff,eq:tau-late,eq:renormtube,eq:renormdis}.
Negligible differences  were found in the NSE time window with the corresponding calculations for $\alpha = 4/3$.
The obtained theoretical curves (lines) are compared with the 
experimental and simulation results in \ref{dyn-short-long} and \ref{dyn-simu-theo}, respectively. The bare undilated tube $d$ was left as a fit parameter.
The values of $d$ obtained for each star are summarized in \ref{tbl:tube}.
The resulting value for the undilated confinement size in all the stars is about 35 $\text{\AA}$.
This value is much smaller than the tube diamater reported for linear polyethylene, $d = 49$ $\text{\AA}$ \cite{Wischnewski2002}. 
However, it must be stressed that both values have been obtained in the framework of two theories (Vilgis-Bou\'{e} model for the PE stars and the standard tube model \cite{Doi1988} for linear PE) that are constructed on different assumptions, so that equivalence of the obtained tube diameters should not be expected.

\begin{table}
  \caption{Effective bare tube diameter from NSE experiments and MD simulations}
  \begin{tabular}{lll}
    \hline
  sample: & NSE & MD  \\
    \hline
 3 arm small &	$35.4 \pm 0.2$ \AA &	 $35.7 \pm 0.2$	\AA \\
 4 arm small &  $34.9 \pm 0.2$ \AA & $34.4 \pm 0.2$ \AA \\
 5 arm small & & $33.2 \pm 0.2$ \AA \\
    \hline
 3 arm large & $35.9 \pm 0.2$ \AA & \\
 4 arm large & $36.2 \pm 0.2$ \AA & \\
						
    \hline
  \end{tabular}
  \label{tbl:tube}
\end{table}


We have also compared the theoretical model with the simulations of the small stars,
see \ref{dyn-simu-theo} for $f = 3$, 4 and 5. 
The fits provide values of $d$ that, after rescaling by $\gamma_q$, are within the error bars of those obtained from the NSE analysis ($d \approx 35$ $\text{\AA}$ see \ref{tbl:tube}). The results have been obtained by fitting the theory to the equivalent NSE time window (up to 400 ns) and extending the calculation for longer times.
The agreement with the theory worsens by increasing the functionality. For the 3-arm stars a very good description is achieved up to the
time limit of validity of DTD, $t^{\ast} = 10^6 \tau_0$. However for $f = 4$ and 5 the theory clearly overestimates the decay at times $t > 3\times 10^5 \tau_0$ and $10^5 \tau_0$ respectively, i.e., much earlier than $t^\ast$. 
Such deviations have not been detected in the NSE experiments. Indeed they only arise at times scales that are beyond the NSE time window 
(dashed line in \ref{dyn-simu-theo}).

The observations in \ref{dyn-simu-theo} are consistent with those anticipated in the MSD (\ref{msd-scaled}). In the three simulated stars with identical $Z_{\rm a} =5$, we find the same amount of relaxed material at the same time, irrespective of the functionality (\ref{tube-surv}). Therefore the theory predicts the same renormalization 
of the tube parameters (thorugh \ref{eq:renormtube} and \ref{eq:renormdis}).
This is in contradiction with the very different relaxation found at long times in the corresponding $S(q,t)$ functions.
Indeed, despite DTD being wrong in itself for $t > t^{\ast} = 10^6\tau_0$, 
the deviations between simulations and theory for $f = 4$ and $f=5$
already arise at times much earlier than $t^{\ast}$ (see \ref{dyn-simu-theo}). In summary, the simulation results,
which reflect the functionality dependence of the relaxation
of the branch point, in spite of the observed independence in the tube survival probability, are in contradiction with predictions from DTD,
and suggest the need of revising current tube models to explicitly account for the number of arms in the mechanisms leading to relaxation.



\section{Conclusions}

We have presented a systematic study, by combining MD simulations and NSE experiments, of symmetric star polymers of different molecular weights and functionalities to investigate the influence of such parameters in the branch point mobility. From the simulation data we have characterized the motion of the branch point through its mean squared displacement, its fluctuations around the mean path and its dynamic structure factor.  By proton labelling of the polyethylene star centers and deuteration of the rest, we have got direct access to the branch point motion through the NSE dynamic structure factor.
The simulations and the experiments reveal a stronger confinement of the branch point by increasing the functionality, following 
the  $2/f$-scaling proposed by Warner \cite{Warner1981}. 
Evidence of deep diving modes has been found only for $f = 3$ within the simulation time scale, indicating that increasing the functionality largely
hinders deep explorations of the arm tubes stemming from the branch point.
The strong localization induced by branching is not limited to the segments around the star center but extends up to a distance of about three entanglements from the branch point. Increasing the functionality does not only result in a stronger localization, but also in a slower relaxation of the branch point at long times. No effect of the functionality is however found within statistics in the tube survival probability of the whole system.

We have analyzed the dynamic structure factors of the branch point in terms 
of a modified version of the model of Vilgis and Bou\'{e} \cite{Vilgis1988} for cross-linked networks, where we have incorporated DTD.
Since DTD renormalization of the tube parameters depends only on the tube survival probability, which does not show any dependence
on $f$, the DTD theory predicts no differences in the relaxation of the branch point in stars with different $f$
and the same arm length.  
However, the simulations, which have accessed time scales much longer than the NSE experiments, reveal a strong $f$-dependence of the branch point relaxation, even at time scales within the expected range of validity of DTD. In particular, the theory overestimates the relaxation, which is slower in the real data. The quality of the theoretical description worsens by increasing the functionality. 
This set of results suggests the need of revising current tube models to explicitly account for the number of arms in the mechanisms leading to relaxation, in particular in the surroundings of the branch point.


\begin{acknowledgement}

We acknowledge financial support from the projects 
MAT2015-63704-P (MINECO-Spain and FEDER-UE) and IT-654-13 (Basque Government, Spain).

\end{acknowledgement}



\providecommand{\latin}[1]{#1}
\providecommand*\mcitethebibliography{\thebibliography}
\csname @ifundefined\endcsname{endmcitethebibliography}
  {\let\endmcitethebibliography\endthebibliography}{}

\newpage

\begin{figure}[htbp!]
\begin{center}
\includegraphics[width=0.5\linewidth]{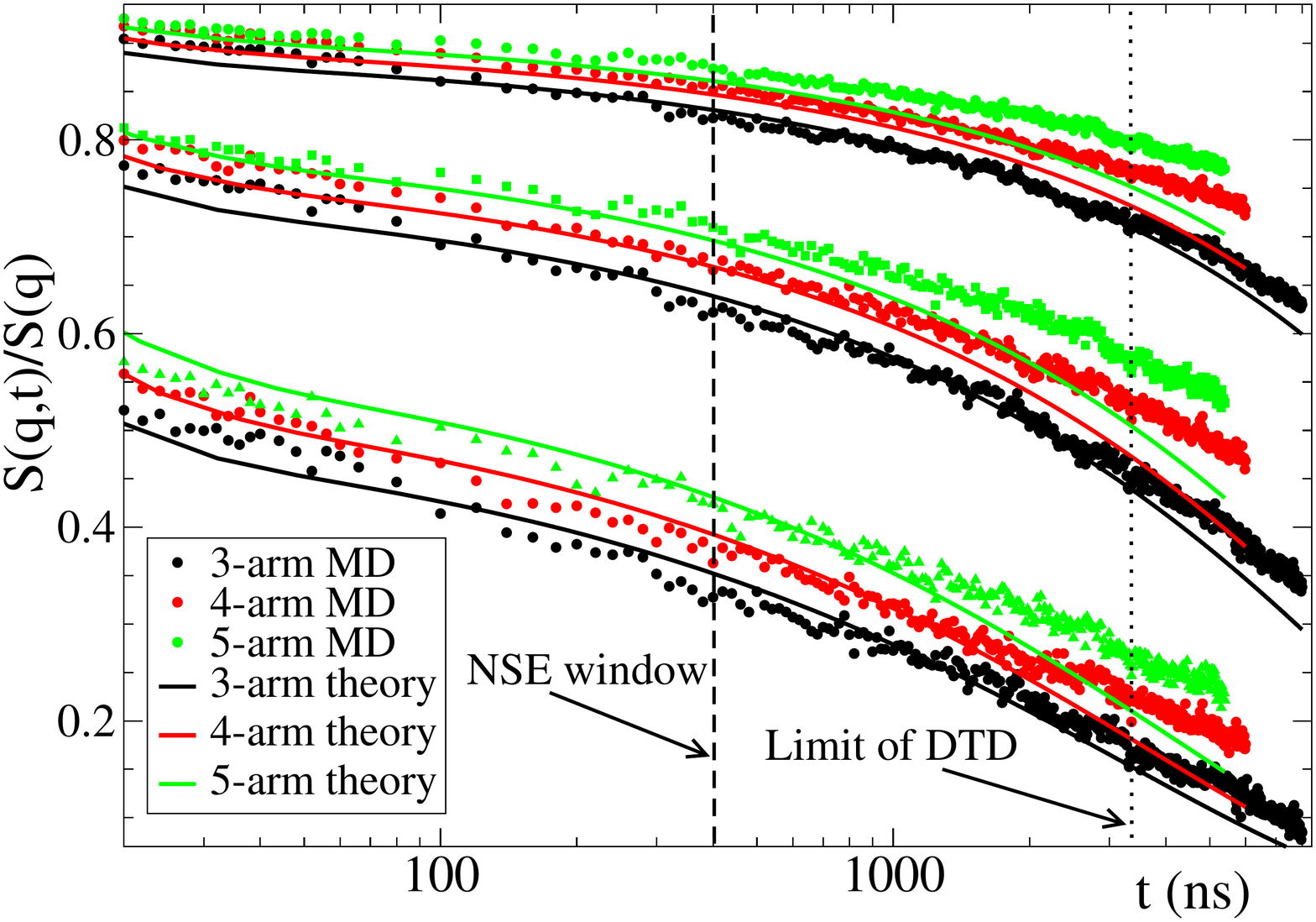}
\end{center}
\end{figure}

\begin{center}
TABLE OF CONTENTS

\end{center}

\end{document}